\documentstyle[pre,aps,amsfonts,amssymb,eqsecnum,amsmath]{revtex}
\begin{document}
\def\phi{\varphi} \def\epsilon{\varepsilon} \def\u{\bbox}
\def\I{{\mathbb I}^-}

\draft \title{Application of renormalization to the dynamics of a
  particle in a infinite square-well potential driven by an external
  field}

\author{C.\ Chandre} \address{Service de Physique Th\'eorique, CEA
  Saclay, F-91191 Gif-sur-Yvette Cedex, France} \maketitle
\date{\today}

\begin{abstract}
  We analyze by a renormalization method, the dynamics of a particle
  in a infinite square-well potential driven by an external
  monochromatic field. This method set up for Hamiltonian systems with
  two degrees of freedom allows us to analyze precisely the stability
  of the trajectories of the particle as a function of the amplitude
  $\varepsilon$ of the external field. We compute numerical values of
  $\varepsilon$ for which the motion of the particle with frequency
  $\omega$ is broken and a transition to a chaotic behavior occurs. 
  We obtain the critical
  function $\varepsilon_c(\omega)$ associated with this system as a
  function of the parameters such as the frequency of the field and
  the width of the potential.
  
\end{abstract}

\section{Introduction}\label{sec.1}

Due to the existence of as many conserved quantities as degrees of
freedom, the trajectories of an integrable Hamiltonian system are
confined to evolve on invariant rotational tori. On a given torus, the dynamics
is regular, i.e.\ conjugate to a linear flow with frequency $\omega$
in action-angle coordinates. This regularity is broken by any
small perturbation, and chaotic trajectories appear~: The phase space
of a Hamiltonian system close to integrable is in general a mixture of
regular and chaotic motions. For Hamiltonian systems with two degrees
of freedom, these invariant rotational tori, also called KAM
(Kolmogorov-Arnold-Moser) tori, act as barriers in phase space.\\
For a
given one parameter family of Hamiltonians $\{H_{\varepsilon}\}$ with
$H_{\varepsilon=0}$ integrable and $\varepsilon$ the amplitude of the
perturbation, it appears from numerical evidences that for
$\varepsilon$ smaller than a critical value denoted
$\varepsilon_c(\omega)$, there exists a KAM torus with frequency
$\omega$, and this torus is broken for larger values by resonance or
overlapping of resonances. The function $\omega \mapsto
\varepsilon_c(\omega)$ is called the critical function associated with
the one-parameter family of Hamiltonians $\{H_{\varepsilon}\}$. This
function contains the information on the existence of invariant tori
in phase space as the parameter $\varepsilon$ increases. In
particular, for $\varepsilon > \sup_{\omega \in B}
\varepsilon_c(\omega)$, there is no longer any KAM torus in the region
of phase space corresponding to the set of frequencies $B$, and we
have large scale stochasticity.

Renormalization methods have been defined and studied numerically for
the analysis of stability of Hamiltonian systems with two degrees of
freedom~\cite{esca81,koch99,govi97,chan98a,chan98b,abad98}.  The aim
is to describe the break-up of a given invariant torus.  The idea is
to set up a transformation that focuses on a specific region of phase
space around the given torus. It acts as a microscope in phase space,
looking at the system on smaller scales in phase space and on longer
time scales.  The complete renormalization
method~\cite{koch99,chan98b,abad98} is a tool to compute precisely the
critical function $\varepsilon_c(\omega)$.  It has been verified that
for specific models (like a forced pendulum) the critical couplings
obtained by renormalization coincide with other methods like Greene's
residue criterion~\cite{gree79} or Laskar's frequency map
analysis~\cite{lask99,chan00a,chan00d}.

The system we analyze is a particle of mass $m$ in an infinite
square-well potential $V_{SQ}$ of width $2a$ driven by an external
monochromatic field with amplitude $\varepsilon$ and frequency
$\Omega$. The Hamiltonian of this system with 1.5 degrees of freedom
is the following~:
\begin{equation}
  \label{eq:Ham1.5}
  H(p,x,t)=\frac{p^2}{2m}+V_{SQ}(x)+\varepsilon x \cos(\Omega t),
\end{equation}
where
$$
V_{SQ}(x)=0 \mbox{ for } |x| < a \mbox{ and } V_{SQ}(x)=+\infty
\mbox{ for } |x|\geq a,
$$
and $\varepsilon \geq 0$.  This system has been studied in
Refs.~\cite{lin86,Breic92} by approximate renormalization methods.
Without external field, i.e.\ for $\varepsilon=0$, the system is
integrable and the motion is periodic with frequency $\omega=
\sqrt{\frac{\pi^2 E}{2ma^2}}$ where $E$ denotes the energy of the
system. For $\varepsilon \not= 0$, some of these regular motions
disappear. In particular, there are resonances when the frequency of
the external field $\Omega$ is commensurate with the frequency of the
motion, i.e.\ when $\Omega$ is equal to $\frac{P}{Q} \omega$ ($P$ and
$Q$ are relatively prime integers).  The interaction of these resonances breaks
up some invariant tori (in the spirit of Chirikov's
criterion~\cite{chir79,esca85}). The critical function
$\varepsilon_c(\omega ;m,a,\Omega)$ is the critical value of the
amplitude of the field for which the motion with frequency $\omega$ is
broken.

The aim of this paper is to apply renormalization to a specific model
and to compute numerically $\varepsilon_c(\omega;m,a,\Omega)$. We use
this critical function to locate chaotic zones and to determine
critical parameters for which large scale stochasticity occurs. We
compare some of the results obtained by the renormalization transformation with
the ones obtained by other existing methods such as
Greene's residue criterion, in order to validate the results obtained
by renormalization.

In Sec.~\ref{sec:2}, we explicit the model and give some
information on its dynamics. In Sec.~\ref{sec:3}, we give a short
description of the renormalization method, and in Sec.~\ref{sec:4} we
compute numerically and analyze the critical function
$\varepsilon_c(\omega;m,a,\Omega)$.


\section{Model}
\label{sec:2}

The external field induces resonances in the system when
$\Omega=\frac{P}{Q} \omega$ where $\omega$ is the frequency of the
unperturbed motion, $\Omega$ the one of the external field and $P$,
$Q$ relatively prime integers. Due to
the specific form of the interaction between the particle and the
field, the largest resonances (i.e.\ of order of the amplitude
$\varepsilon$ of the field) are obtained with $Q=1$ and $P$ odd. This
can be seen by writing Hamiltonian~(\ref{eq:Ham1.5}) in action-angle
variables~\cite{Breic92}~:
\begin{equation}
  \label{eq:Ham-aa}
  H(A,\phi,t)=\frac{\pi^2}{8ma^2} A^2 -\frac{4\varepsilon a}{\pi^2} 
  \sum_{\scriptstyle n \in
  {\mathbb{Z}}\atop n \, \mathrm{odd}} 
\frac{1}{n^2} \cos(n\phi -\Omega t).
\end{equation}
There is a resonance when $n\dot{\varphi} =\Omega$, which corresponds
to $n\omega=\Omega$ since $\dot{\varphi}=\frac{\partial H}{\partial
  A}$ is the frequency $\omega$ of the motion. This resonance will be
denoted 1:$n$ in what follows.  Hamiltonian~(\ref{eq:Ham-aa}) can be
mapped into a time-independent Hamiltonian with two degrees of
freedom by considering $-\Omega t$ as a new angle variable~:
\begin{equation}
  \label{eq:Ham2}
  H(A_1,A_2,\phi_1,\phi_2)=\frac{\pi^2}{8ma^2}A_1^2 -\Omega A_2 -
\frac{4\varepsilon a}{\pi^2} \sum_{n\, \mathrm{
  odd}} \frac{1}{n^2} \cos(n\phi_1 +\phi_2).
\end{equation}
We rescale time by a factor $\Omega$, i.e.\ we multiply
Hamiltonian~(\ref{eq:Ham2}) by $1/\Omega$. We notice that this
rescaling of time changes the frequency of a quasiperiodic motion of
Hamiltonian~(\ref{eq:Ham2}) by a factor $\Omega$. The new rescaled
frequency is now $\omega/\Omega$. We rescale the action
variables by replacing $H(\u{A},\u{\phi})$ by
$\lambda^{-1}H(\u{A}/\lambda,\u{\phi})$ with a factor
$\lambda=\pi^2/(4ma^2 \Omega)$. We notice that the rescaling in the
actions does not change the equations of motion.  After this
rescaling, Hamiltonian~(\ref{eq:Ham2}) is equal to
\begin{equation}
  \label{eq:Ham-aa2}
  H(\u{A},\u{\phi})=\frac{A_1^2}{2}-A_2 -\varepsilon'  \sum_{n \, \mathrm{
  odd}} \frac{1}{n^2} \cos(n\phi_1 +\phi_2),
\end{equation}
where $\varepsilon'$ is the \textit{dimensionless} 
amplitude of the external field given by~:
\begin{equation}
  \label{eq:epsdl}
  \varepsilon'=\frac{\varepsilon}{ma\Omega^2}.
\end{equation}
For $\varepsilon' =0$, Hamiltonian~(\ref{eq:Ham-aa2}) depends only on
$\u{A}=(A_1,A_2)$ and the equations of motion show that $A_1(t)$ and
$A_2(t)$ are constant, and $\varphi_1(t)=\omega t+\varphi_{1,0}$ and
$\varphi_2(t)=-t+\varphi_{2,0}$.  The trajectories of the
system~(\ref{eq:Ham-aa2}) evolve on three-dimensional energy surfaces
in the four-dimensional phase space, and for $\varepsilon'=0$ the
trajectories with frequency $\omega$ are confined to evolve on a
two-dimensional torus (on the energy surface) with frequency vector
$\u{\omega}=(\omega,-1)$. \\
For $\varepsilon' >0$, this system has an infinite number of main
resonances (given by the condition $n\dot{\phi_1}+\dot{\phi_2}\approx
0$) located around $A_1=1/n$ where $n$ is an odd integer, accumulating
at $A_1=0$.  The width of the $n$-th resonance zone is approximately
equal to $4\sqrt{\varepsilon'}/n$~\cite{lin86}. For large values of
$\varepsilon'$ the torus with frequency vector $\u{\omega}$ is broken
by overlapping of resonances. In order to have an estimate of the
critical value of $\varepsilon'$ of the break-up of the invariant
torus with frequency $\omega$, we apply Chirikov's
criterion~\cite{chir79}. For a torus with frequency $\omega$ located
between the two primary resonances 1:$n$ and 1:$n$+2, the overlapping
is obtained when the sum of the two half-width of these two resonances
is equal to their distance, i.e.\ for
\begin{equation}
  \label{eq:chirikov}
  \varepsilon^{(c)}=\frac{1}{4(n+1)^2}.
\end{equation}
This value overestimates the critical values of the threshold of the
break-up as it has been noticed in Ref.~\cite{lin86} for this model.
A convenient way to compute the value of $\varepsilon'$ for which the
torus is broken is to set up a complete
renormalization~\cite{koch99,chan98b,abad98,chan00a} in the spirit of
Refs.~\cite{esca81,esca85}.


\section{Renormalization method}
\label{sec:3}

In this section, we give a description of the renormalization method
we apply to the model described in the previous section.

First, we shift the actions such that the invariant torus with
frequency $\omega$ is located at $A_1=0$ for the unperturbed
Hamiltonian (for $\varepsilon'=0$)~: $A'_1=A_1-\omega$ and $A'_2=A_2$.
Hamiltonian~(\ref{eq:Ham-aa2}) is equal to~:
\begin{equation}
  \label{eq:Hstart}
  H(\u{A}',\u{\phi})=\u{\omega}\cdot\u{A}'+\frac{1}{2} 
  (\u{\Omega}\cdot\u{A}')^2
  -\varepsilon'\sum_{\scriptstyle \u{\nu}=(n,1)\atop n \, \mathrm{ odd}} 
 \frac{1}{n^2}
\cos(\u{\nu}\cdot\u{\varphi} ),
\end{equation}
where $\u{\Omega}=(1,0)$ and $\u{\omega}=(\omega,-1)$.\\
The renormalization transformation is a map $H'={\mathcal{R}} (H)$
acting within the family of Hamiltonians of the form~:
\begin{equation}
H({\u A},{\u \varphi})=\u{\omega}\cdot\u{A}
+V({\u \Omega}\cdot{\u A},{\u \varphi}),
\end{equation}
and Hamiltonian~(\ref{eq:Hstart}) is an element of this family.
Furthermore Hamiltonian~(\ref{eq:Hstart}) is quadratic in the actions;
this feature is useful for a simplification of the implementation of
the transformation (see Step 4 below)~\cite{chan98b}.  However, a
similar version of the transformation (obtained by slightly changing
the way Step 4 is
implemented) can be defined and studied numerically~\cite{abad98,chan00d}.\\
The transformation ${\mathcal R}$ is based on the continued fraction
expansion of $\omega$~:
$$
\omega =\frac{1}{a_0+\frac{1}{a_1+\cdots}} \equiv [a_0,a_1,\ldots].
$$
The best rational approximates of $\omega$ are given by the
truncations of its continued fraction expansion~:
$$
\frac{p_k}{q_k}=[a_0,a_1,\ldots,a_k=\infty].
$$
The corresponding periodic orbits with frequency vectors
$(p_k/q_k,-1)$ which are orthogonal to the modes ${\u \nu}_k=(q_k,p_k)$ (${\u
  \nu}_k$ is called ``resonance'' and is also denoted $p_k$:$q_k$ in
what follows) accumulate at the invariant torus with frequency vector
${\u \omega}=(\omega,-1)$.  This family of periodic orbits satisfies
the following relations~: $|{\u \omega}\cdot{\u \nu}_{k+1}| < |{\u
  \omega}\cdot{\u \nu}_k|$ and $ \lim_{k\to \infty} |{\u
  \omega}\cdot{\u \nu}_k|=0$, and $\u{\nu}_k$ is given by the
following equation~:
\begin{equation}
\label{eqn:App2nuk}
{\u \nu}_k=N_{a_0}\cdots N_{a_{k-1}}{\u \nu}_0,
\end{equation}
where ${\u \nu}_0=(1,0)$ and $N_{a_i}$ denotes the matrix
$$
N_{a_i}=\left( \begin{array}{cc} a_i & 1 \\ 1 & 0 \end{array}
\right).$$
The set of the Fourier modes of the perturbation with
frequency vectors $\u{\nu}_k$ leads to the divergence of perturbation
expansions since the small denominators which are equal to
$\u{\omega}\cdot\u{\nu}_k$ tend to zero as $k$ increases. The
renormalization transformation deals with the modes $\u{\nu}_k$
specifically by non-perturbative techniques.

The transformation ${\mathcal R}$ is composed by four steps~:\\
\indent {\bf (1)} A shift of resonances constructed from the condition
${\u \nu}_1 \mapsto {\u \nu}_0$~: we impose that $\cos[(a_0,1)\cdot
{\u \phi}]=\cos [(1,0)\cdot{\u \phi}']$ where $a_0=[\omega^{-1}]$ is
the integer part of $\omega^{-1}$. This change of coordinates is
performed by a linear canonical transformation
$$({\u A},{\u \phi})\mapsto ({\u A}',{\u \varphi}')=(N_{a_0}^{-1}{\u
  A},N_{a_0}{\u \phi}),$$
which is generated by
$F(\u{A}',\u{\varphi})=N_{a_0}\u{A}'\cdot\u{\varphi}$.  The
Hamiltonian expressed in the new coordinates becomes
\begin{eqnarray*}
H'(\u{A}',\u{\varphi}')&=&H(\u{A},\u{\varphi})=\u{\omega}\cdot\u{A}
+V({\u \Omega}\cdot{\u A},{\u \varphi}),\\
&=& \u{\omega}\cdot N_{a_0}\u{A}'
+V({\u \Omega}\cdot N_{a_0}{\u A}',N_{a_0}^{-1}{\u \varphi}'),\\
&=& N_{a_0}\u{\omega}\cdot\u{A}'
+V(N_{a_0}{\u \Omega}\cdot{\u A}',N_{a_0}^{-1}{\u \varphi}').
\end{eqnarray*}
Thus the new frequency vector is equal to
$N_{a_0}\u{\omega}=-\omega\u{\omega}'$ where
$\u{\omega}'=(\omega',-1)$ and $\omega'$ is the image of $\omega$ by
the Gauss map
\begin{equation}
\label{eqn:gauss}
\omega\mapsto \omega'=\omega^{-1}-[\omega^{-1}].
\end{equation} 
This map corresponds to a shift to the left of the entries in the
continued fraction expansion of the frequency
$$
\omega=[a_0,a_1,a_2,\ldots]\mapsto \omega'=[a_1,a_2,a_3,\ldots].
$$
The main effect of Step 1 is to change the frequency vectors of the
Fourier modes of the perturbation according to the map $\u{\nu}
\mapsto N_{a_0}^{-1}\u{\nu}$.

\indent {\bf (2)} We rescale the energy by a factor $\omega^{-1}$ (or
equivalently time by a factor $\omega$), i.e.\ we multiply the
Hamiltonian by $\omega^{-1}$, and we change the sign of
both phase space coordinates $({\u A},{\u \phi})\mapsto (-{\u A},-{\u
  \phi})$, in order to have ${\u \omega}'$ as the new frequency
vector, i.e.\ such that the linear term in the actions
$N_{a_0}\u{\omega}\cdot \u{A}=-\omega\u{\omega}'\cdot\u{A}$ is
rescaled into a term of the form ${\u \omega}'\cdot{\u A}$. Moreover,
${\u \Omega}=(1,\alpha)$ is changed into
$N_{a_0}\u{\Omega}=(a_0+\alpha,1)=(a_0+\alpha)\u{\Omega}'$ with ${\u
  \Omega}'=(1,\alpha')=(1,(a_0+\alpha)^{-1})$ since the normalization
condition we use is that the first component of $\u{\Omega}$ is equal
to one.  The map $\alpha \mapsto (a_0+\alpha)^{-1}$ is the inverse of the
Gauss map~(\ref{eqn:gauss}), in the sense that if
$\alpha=[b_0,b_1,\ldots]$ then $\alpha'=[a_0,b_0,b_1,\ldots]$.  If we
define
$$
[\alpha|\omega]=[\ldots,b_2,b_1,b_0|a_0,a_1,a_2,\ldots],
$$
The map $[\alpha|\omega]\mapsto [\alpha'|\omega']$ corresponds to a
two-sided Bernoulli shift~:
$$
[\ldots,b_2,b_1,b_0|a_0,a_1,a_2,\ldots] \mapsto
[\ldots,b_2,b_1,b_0,a_0|a_1,a_2,\ldots].
$$
Since $\omega <1$, the effect of the rescaling of time is that the
trajectories of the rescaled Hamiltonian correspond to the ones of the
initial Hamiltonian on a longer time scale.

\indent {\bf (3)} We perform a rescaling of the actions~: $H$ is
changed into
$$H'({\u A},{\u \phi})=\lambda H\left(\frac{{\u A}}{\lambda},{\u \phi}
\right),$$
with $\lambda=\lambda(H)$ such that the mean value of the
quadratic term of $H'$ in the variable ${\u \Omega}'\cdot{\u A}$, is
equal to $(\u{\Omega}'\cdot\u{A})^2/2$~:
\begin{equation}
\label{eqn:appli2dlamb}
\lambda=2\omega^{-1}(a_0+\alpha)^2\langle V^{(2)} \rangle,
\end{equation}
where $V^{(2)}$ denotes the coefficient of the quadratic term of
$V(\u{\Omega}\cdot\u{A},\u{\varphi})=H-\u{\omega}\cdot \u{A}$ in the
variable $\u{\Omega}\cdot \u{A}$, and $\langle V^{(2)} \rangle$
denotes its mean value on $[0,2\pi[^2$. Since in general $\langle
V^{(2)} \rangle$ is close to $1/2$, the rescaling
coefficient~(\ref{eqn:appli2dlamb}) is larger than one. Thus the
rescaling in the actions corresponds to a focus in phase space around
the invariant torus with frequency $\u{\omega}'$ located approximately
at $\u{\Omega}'\cdot\u{A} \approx 0$.

After these three steps, $H$ is changed into
$$
H'({\u A},{\u
  \phi})=\u{\omega}'\cdot\u{A}+2\omega^{-2}(a_0+\alpha)^2 \langle
V^{(2)} \rangle V\left(- \frac{1}{2\omega^{-1}(a_0+\alpha)\langle
    V^{(2)} \rangle} \u{\Omega}'\cdot{\u A},-N_{a_0}^{-1} {\u
    \varphi}\right).
$$
\indent {\bf (4)} The fourth step is a canonical transformation
that eliminates the non-resonant part (denoted $I^-$)
of the perturbation of $H'$. \\
The choice of the part of the perturbation which has to be considered
resonant or non-resonant is somewhat arbitrary. The set of non-resonant modes contains
the modes of the perturbation which are sufficiently far from the
resonances $\{\u{\nu}_k\}$ in order to avoid small denominator
problems during the elimination process. A convenient choice
concerning the non-resonant modes is the set $I^-$ of integer vectors
${\u \nu}=(\nu_1,\nu_2)$ such that $|\nu_2|> |\nu_1|$~:
$$
I^-=\{ {\u \nu}=(\nu_1,\nu_2)\in{\Bbb Z}^2 \, | \, |\nu_2|> |\nu_1|
\}.
$$
We notice that Eq.~(\ref{eqn:App2nuk}) defining ${\u
  \nu}_k=(q_k,p_k)$ shows that $p_k\leq q_k$ for $k\geq 0$.
Consequently, the resonances do not belong to $I^-$.  At each
iteration of the transformation, the frequency vector of the
considered torus changes (since we perform unimodular
transformations). We have chosen a unique region $I^-$ such that it
does not contain any of the resonance lines ${\u \omega}\cdot {\u \nu}
=0$ for all
$\omega \in ]0,1[$.\\
From the form of the eigenvectors of $N_{a_i}$, we can see that each
vector ${\u \nu}\in {\mathbb Z}^2\setminus \{{\u 0}\}$ is mapped into
$I^-$ after a sufficient number of iterations of the matrices
$N_{a_i}$ (the eigenvector of $N_{a_i}^{-1}$ associated with the
eigenvalue of modulus larger than one points into $I^-$).  In other
terms, each resonant mode becomes non-resonant at a sufficiently
smaller scale in phase space. We notice
that ${\u 0}$ is not an element of $I^-$, i.e.\ it is resonant.\\
Since the initial Hamiltonian~(\ref{eq:Ham-aa2}) is quadratic in the
actions, the renormalization for quadratic Hamiltonians defined in
Refs.~\cite{chan98b,chan00a} is well-suited for this problem.  We
define more precisely Step 4 for the following Hamiltonians~:
$$
H({\u A},{\u \phi})={\u \omega}\cdot{\u A}+ m({\u \phi})({\u
  \Omega}\cdot{\u A})^2 + g({\u \phi}){\u \Omega} \cdot {\u A} +f({\u
  \phi}),
$$
where $m$, $g$ and $f$ are scalar functions of the angles and
$\langle m \rangle \not= 0$. We eliminate completely the non-resonant
modes of $g$ and $f$ by a canonical transformation connected to the
identity, which is defined by iterating KAM-type transformations.  The
KAM iterations we perform (by Lie transformations) are generated by
functions that are linear in the actions. The Hamiltonian expressed in
the new coordinates is again quadratic in the actions. Thus this type
of transformation allows us to remain quadratic at each step of the
transformation~\cite{Bthir92}. One iteration ${\mathcal U}_S$ of the
KAM transformation reduces the non-resonant modes of $f$ and $g$, from
order $\varepsilon$ to $\varepsilon^2$. The transformation that
eliminates completely the non-resonant part is defined in the
following way~:
$$
H'=H\circ {\mathcal U}_H, \, \mbox{ where } {\mathcal U}_H=
{\mathcal U}_{S_1}\circ {\mathcal U}_{S_2} \circ \cdots {\mathcal
  U}_{S_n}\circ \cdots,
$$
where the purpose of ${\mathcal U}_{S_n}$ is to reduce the
non-resonant part of $f$ and $g$ from order $\varepsilon^{2^{n-1}}$ to
$\varepsilon^{2^{n}}$, such that ${\I} f'={\I} g'=0$, where $\I f'$
denotes the non-resonant part of the constant term in the actions of
$H'$, i.e.\ ${\I } f'=\sum_{\u{\nu}\in I^-} f'_{\u{\nu}}
e^{i\u{\nu}\cdot \u{\varphi}}$. For the explicit equations of this
part of the transformation, we refer to Refs.~\cite{chan98b,chan99c}.

In summary, the renormalization transformation acts in the following
way~: First, some of the resonant modes are turned non-resonant by a
rescaling of phase space that changes the frequency of the torus
according to the Gauss map~(\ref{eqn:gauss}). Then an iteration of a
KAM-type transformation eliminates the non-resonant modes (by slightly
changing the resonant ones).\\
The renormalization transformation is a map of Fourier coefficients~:
If
$$
H(\u{A},\u{\varphi} ) =\u{ \omega}  \cdot \u{A} +  
\sum\limits_{k,\u{\nu} } {H_{k,\u{\nu} } (\u{\Omega}  
\cdot \u{A})^k e^{i\u{\nu}  \cdot \u{\varphi} } } ,
$$
then $H'=\mathcal{R}(H)$ becomes $
\left\{ {H_{k,\u{\nu} } '  } \right\} = {\mathcal{R}}\left( {\left\{ {H_{k,\u{\nu} } } \right\}} \right)
$ where $k=0,1,2$ in our case. The properties of $\mathcal{R}$ are obtained by iterating this
map. In order to analyze it numerically, we truncate this map, i.e.\
we truncate the Fourier series by considering frequency vectors
$\u{\nu}=(\nu_1,\nu_2)$ such that $
\max _{i = 1,2} \left| {\nu _i } \right| \leq L$. We observe the
convergence of the properties of the renormalization transformation as
$L$ grows.


\section{Critical function of the model}
\label{sec:4}

\subsection{Critical function $\varepsilon_c'(\omega)$}
\label{sec:41}

The aim is to obtain the value $\varepsilon'_c(\omega)$ for which the
invariant torus with frequency $\omega$ exists for
$\varepsilon'<\varepsilon'_c(\omega)$ and is broken for larger values.

If $\omega$ satisfies a Diophantine condition, the KAM
theorem~\cite{gall83} ensures the persistence of an invariant torus
with frequency $\omega$ for the system we consider, i.e.\ 
$\varepsilon'_c(\omega) > 0$. If $\omega$ is rational, a resonance
breaks up the torus, i.e.\ $\varepsilon'_c(\omega)=0$. Moreover, this
function is symmetric~:
$\varepsilon'_c(-\omega)=\varepsilon_c'(\omega)$. This symmetry comes
from the fact that the canonical transformation
$(A_1,\varphi_1,A_2,\varphi_2)\mapsto (-A_1,-\varphi_1,A_2,\varphi_2)$
only changes the frequency $\omega$ into $-\omega$ in
Hamiltonian~(\ref{eq:Ham-aa2}). \\
The critical function $\varepsilon'_c(\omega)$ is determined by
looking at the iterates of the renormalization transformation
$\mathcal{R}$ described above, i.e.\ it is defined by the following
equations~:
\begin{eqnarray}
&& {\mathcal R}^nH_{\varepsilon'} \underset{n\to \infty}{\to}
H_0(\u{A}) =\u{\omega}\cdot\u{A}+
\frac{1}{2}(\u{\Omega}\cdot\u{A})^2 \quad \mbox{ for }
\varepsilon'<\varepsilon'_c(\omega), \label{eqn:Adef1}\\
&& {\mathcal R}^nH_{\varepsilon'} \underset{ n\to \infty}{\to} \infty \quad \mbox{ for }
\varepsilon'>\varepsilon'_c(\omega),\label{eqn:Adef2}
\end{eqnarray}
where $H_{\varepsilon'}$ is Hamiltonian~(\ref{eq:Hstart}).
Numerical work shows that Eqs.(\ref{eqn:Adef1})-(\ref{eqn:Adef2})
define uniquely $\varepsilon'_c(\omega)$, i.e.\ the renormalization
has two main domains~: one where the iterations of the transformation
converge to an integrable Hamiltonian $H_0$ --- this domain is
conjectured to be the set of Hamiltonians that have a smooth invariant
torus with frequency $\omega$ --- and a domain where the iterations
diverge, and this domain is conjectured to be the set of Hamiltonians
that do not have an invariant torus with the frequency $\omega$. For
$\varepsilon'=\varepsilon'_c(\omega)$, it was shown numerically that
the iterations converge to a strange chaotic attractor containing all
the relevant information on critical tori (i.e.\ at the threshold of
the break-up)~\cite{chan00a}. This means that we conjecture that the critical
thresholds obtained by the complete renormalization coincide (up to
numerical precision) with the thresholds of the break-up of invariant
tori. In order to give support to this conjecture, we compare the
values we obtain with another independent method, Greene's residue
criterion. This method consists in analyzing the stability of nearby
periodic orbits (for instance, the ones given by the truncations of
the continued fraction expansion of the frequency of the torus). For
rigorous results on Greene's criterion, we refer to
Refs.~\cite{mack92,falc92}.  

\begin{figure}
\caption{\label{fig:epsc}Critical function of Hamiltonian system~(\ref{eq:Hstart})}

\vspace*{1cm}

\setlength{\unitlength}{0.240900pt}
\ifx\plotpoint\undefined\newsavebox{\plotpoint}\fi
\sbox{\plotpoint}{\rule[-0.200pt]{0.400pt}{0.400pt}}%
\begin{picture}(1500,900)(0,0)
\font\gnuplot=cmr10 at 10pt
\gnuplot
\sbox{\plotpoint}{\rule[-0.200pt]{0.400pt}{0.400pt}}%
\put(160.0,82.0){\rule[-0.200pt]{4.818pt}{0.400pt}}
\put(1419.0,82.0){\rule[-0.200pt]{4.818pt}{0.400pt}}
\put(160.0,212.0){\rule[-0.200pt]{4.818pt}{0.400pt}}
\put(140,212){\makebox(0,0)[r]{0.01}}
\put(1419.0,212.0){\rule[-0.200pt]{4.818pt}{0.400pt}}
\put(160.0,341.0){\rule[-0.200pt]{4.818pt}{0.400pt}}
\put(1419.0,341.0){\rule[-0.200pt]{4.818pt}{0.400pt}}
\put(160.0,471.0){\rule[-0.200pt]{4.818pt}{0.400pt}}
\put(140,471){\makebox(0,0)[r]{0.02}}
\put(30,471){\makebox(0,0)[r]{$\varepsilon'_c$}}
\put(1419.0,471.0){\rule[-0.200pt]{4.818pt}{0.400pt}}
\put(160.0,601.0){\rule[-0.200pt]{4.818pt}{0.400pt}}
\put(1419.0,601.0){\rule[-0.200pt]{4.818pt}{0.400pt}}
\put(160.0,730.0){\rule[-0.200pt]{4.818pt}{0.400pt}}
\put(140,730){\makebox(0,0)[r]{0.03}}
\put(1419.0,730.0){\rule[-0.200pt]{4.818pt}{0.400pt}}
\put(160.0,860.0){\rule[-0.200pt]{4.818pt}{0.400pt}}
\put(1419.0,860.0){\rule[-0.200pt]{4.818pt}{0.400pt}}
\put(160.0,82.0){\rule[-0.200pt]{0.400pt}{4.818pt}}
\put(160,41){\makebox(0,0){0.2}}
\put(160.0,840.0){\rule[-0.200pt]{0.400pt}{4.818pt}}
\put(320.0,82.0){\rule[-0.200pt]{0.400pt}{4.818pt}}
\put(320,41){\makebox(0,0){0.3}}
\put(320.0,840.0){\rule[-0.200pt]{0.400pt}{4.818pt}}
\put(480.0,82.0){\rule[-0.200pt]{0.400pt}{4.818pt}}
\put(480,41){\makebox(0,0){0.4}}
\put(480.0,840.0){\rule[-0.200pt]{0.400pt}{4.818pt}}
\put(640.0,82.0){\rule[-0.200pt]{0.400pt}{4.818pt}}
\put(640,41){\makebox(0,0){0.5}}
\put(640.0,840.0){\rule[-0.200pt]{0.400pt}{4.818pt}}
\put(799.0,82.0){\rule[-0.200pt]{0.400pt}{4.818pt}}
\put(799,41){\makebox(0,0){0.6}}
\put(799,-20){\makebox(0,0){$\omega$}}
\put(799.0,840.0){\rule[-0.200pt]{0.400pt}{4.818pt}}
\put(959.0,82.0){\rule[-0.200pt]{0.400pt}{4.818pt}}
\put(959,41){\makebox(0,0){0.7}}
\put(959.0,840.0){\rule[-0.200pt]{0.400pt}{4.818pt}}
\put(1119.0,82.0){\rule[-0.200pt]{0.400pt}{4.818pt}}
\put(1119,41){\makebox(0,0){0.8}}
\put(1119.0,840.0){\rule[-0.200pt]{0.400pt}{4.818pt}}
\put(1279.0,82.0){\rule[-0.200pt]{0.400pt}{4.818pt}}
\put(1279,41){\makebox(0,0){0.9}}
\put(1279.0,840.0){\rule[-0.200pt]{0.400pt}{4.818pt}}
\put(1439.0,82.0){\rule[-0.200pt]{0.400pt}{4.818pt}}
\put(1439,41){\makebox(0,0){1}}
\put(1439.0,840.0){\rule[-0.200pt]{0.400pt}{4.818pt}}
\put(160.0,82.0){\rule[-0.200pt]{308.111pt}{0.400pt}}
\put(1439.0,82.0){\rule[-0.200pt]{0.400pt}{187.420pt}}
\put(160.0,860.0){\rule[-0.200pt]{308.111pt}{0.400pt}}
\put(160.0,82.0){\rule[-0.200pt]{0.400pt}{187.420pt}}
\put(492,325){\rule{1pt}{1pt}}
\put(1052,736){\rule{1pt}{1pt}}
\put(897,554){\rule{1pt}{1pt}}
\put(1242,202){\rule{1pt}{1pt}}
\put(1296,86){\rule{1pt}{1pt}}
\put(1066,754){\rule{1pt}{1pt}}
\put(1260,159){\rule{1pt}{1pt}}
\put(1098,637){\rule{1pt}{1pt}}
\put(1062,680){\rule{1pt}{1pt}}
\put(883,608){\rule{1pt}{1pt}}
\put(287,111){\rule{1pt}{1pt}}
\put(701,358){\rule{1pt}{1pt}}
\put(936,785){\rule{1pt}{1pt}}
\put(1245,198){\rule{1pt}{1pt}}
\put(700,495){\rule{1pt}{1pt}}
\put(490,381){\rule{1pt}{1pt}}
\put(1059,744){\rule{1pt}{1pt}}
\put(763,711){\rule{1pt}{1pt}}
\put(947,509){\rule{1pt}{1pt}}
\put(1116,562){\rule{1pt}{1pt}}
\put(1199,314){\rule{1pt}{1pt}}
\put(284,118){\rule{1pt}{1pt}}
\put(620,273){\rule{1pt}{1pt}}
\put(922,751){\rule{1pt}{1pt}}
\put(693,578){\rule{1pt}{1pt}}
\put(664,359){\rule{1pt}{1pt}}
\put(766,512){\rule{1pt}{1pt}}
\put(786,488){\rule{1pt}{1pt}}
\put(1143,494){\rule{1pt}{1pt}}
\put(848,765){\rule{1pt}{1pt}}
\put(1269,142){\rule{1pt}{1pt}}
\put(711,657){\rule{1pt}{1pt}}
\put(437,206){\rule{1pt}{1pt}}
\put(869,796){\rule{1pt}{1pt}}
\put(312,95){\rule{1pt}{1pt}}
\put(444,209){\rule{1pt}{1pt}}
\put(1262,156){\rule{1pt}{1pt}}
\put(1048,698){\rule{1pt}{1pt}}
\put(559,489){\rule{1pt}{1pt}}
\put(1168,400){\rule{1pt}{1pt}}
\put(939,777){\rule{1pt}{1pt}}
\put(1083,395){\rule{1pt}{1pt}}
\put(729,292){\rule{1pt}{1pt}}
\put(893,690){\rule{1pt}{1pt}}
\put(844,751){\rule{1pt}{1pt}}
\put(559,485){\rule{1pt}{1pt}}
\put(1116,562){\rule{1pt}{1pt}}
\put(256,114){\rule{1pt}{1pt}}
\put(1144,485){\rule{1pt}{1pt}}
\put(726,406){\rule{1pt}{1pt}}
\put(879,755){\rule{1pt}{1pt}}
\put(544,339){\rule{1pt}{1pt}}
\put(529,166){\rule{1pt}{1pt}}
\put(924,761){\rule{1pt}{1pt}}
\put(1293,90){\rule{1pt}{1pt}}
\put(489,357){\rule{1pt}{1pt}}
\put(483,247){\rule{1pt}{1pt}}
\put(1202,299){\rule{1pt}{1pt}}
\put(932,772){\rule{1pt}{1pt}}
\put(1209,282){\rule{1pt}{1pt}}
\put(1034,608){\rule{1pt}{1pt}}
\put(985,440){\rule{1pt}{1pt}}
\put(317,103){\rule{1pt}{1pt}}
\put(745,713){\rule{1pt}{1pt}}
\put(958,797){\rule{1pt}{1pt}}
\put(976,587){\rule{1pt}{1pt}}
\put(433,182){\rule{1pt}{1pt}}
\put(678,385){\rule{1pt}{1pt}}
\put(856,567){\rule{1pt}{1pt}}
\put(1072,722){\rule{1pt}{1pt}}
\put(1008,798){\rule{1pt}{1pt}}
\put(491,351){\rule{1pt}{1pt}}
\put(981,342){\rule{1pt}{1pt}}
\put(1063,610){\rule{1pt}{1pt}}
\put(1212,284){\rule{1pt}{1pt}}
\put(777,656){\rule{1pt}{1pt}}
\put(675,505){\rule{1pt}{1pt}}
\put(793,294){\rule{1pt}{1pt}}
\put(891,689){\rule{1pt}{1pt}}
\put(701,360){\rule{1pt}{1pt}}
\put(493,393){\rule{1pt}{1pt}}
\put(522,170){\rule{1pt}{1pt}}
\put(620,268){\rule{1pt}{1pt}}
\put(454,200){\rule{1pt}{1pt}}
\put(706,653){\rule{1pt}{1pt}}
\put(1269,140){\rule{1pt}{1pt}}
\put(951,792){\rule{1pt}{1pt}}
\put(290,128){\rule{1pt}{1pt}}
\put(1101,628){\rule{1pt}{1pt}}
\put(475,294){\rule{1pt}{1pt}}
\put(698,594){\rule{1pt}{1pt}}
\put(1151,469){\rule{1pt}{1pt}}
\put(795,235){\rule{1pt}{1pt}}
\put(431,167){\rule{1pt}{1pt}}
\put(421,95){\rule{1pt}{1pt}}
\put(1032,711){\rule{1pt}{1pt}}
\put(619,278){\rule{1pt}{1pt}}
\put(1252,181){\rule{1pt}{1pt}}
\put(869,799){\rule{1pt}{1pt}}
\put(772,625){\rule{1pt}{1pt}}
\put(1074,726){\rule{1pt}{1pt}}
\put(282,118){\rule{1pt}{1pt}}
\put(472,265){\rule{1pt}{1pt}}
\put(758,710){\rule{1pt}{1pt}}
\put(227,95){\rule{1pt}{1pt}}
\put(495,380){\rule{1pt}{1pt}}
\put(996,772){\rule{1pt}{1pt}}
\put(252,85){\rule{1pt}{1pt}}
\put(617,313){\rule{1pt}{1pt}}
\put(660,290){\rule{1pt}{1pt}}
\put(1200,299){\rule{1pt}{1pt}}
\put(674,515){\rule{1pt}{1pt}}
\put(659,281){\rule{1pt}{1pt}}
\put(780,643){\rule{1pt}{1pt}}
\put(1283,111){\rule{1pt}{1pt}}
\put(936,799){\rule{1pt}{1pt}}
\put(544,310){\rule{1pt}{1pt}}
\put(1074,634){\rule{1pt}{1pt}}
\put(1166,415){\rule{1pt}{1pt}}
\put(286,122){\rule{1pt}{1pt}}
\put(1008,784){\rule{1pt}{1pt}}
\put(1200,307){\rule{1pt}{1pt}}
\put(666,377){\rule{1pt}{1pt}}
\put(924,713){\rule{1pt}{1pt}}
\put(1079,659){\rule{1pt}{1pt}}
\put(427,134){\rule{1pt}{1pt}}
\put(868,745){\rule{1pt}{1pt}}
\put(659,284){\rule{1pt}{1pt}}
\put(522,168){\rule{1pt}{1pt}}
\put(700,412){\rule{1pt}{1pt}}
\put(1248,190){\rule{1pt}{1pt}}
\put(775,651){\rule{1pt}{1pt}}
\put(663,326){\rule{1pt}{1pt}}
\put(812,485){\rule{1pt}{1pt}}
\put(461,313){\rule{1pt}{1pt}}
\put(941,808){\rule{1pt}{1pt}}
\put(1263,155){\rule{1pt}{1pt}}
\put(998,739){\rule{1pt}{1pt}}
\put(1080,608){\rule{1pt}{1pt}}
\put(1066,758){\rule{1pt}{1pt}}
\put(666,380){\rule{1pt}{1pt}}
\put(620,270){\rule{1pt}{1pt}}
\put(227,87){\rule{1pt}{1pt}}
\put(852,760){\rule{1pt}{1pt}}
\put(251,102){\rule{1pt}{1pt}}
\put(564,404){\rule{1pt}{1pt}}
\put(832,751){\rule{1pt}{1pt}}
\put(470,275){\rule{1pt}{1pt}}
\put(1077,714){\rule{1pt}{1pt}}
\put(923,630){\rule{1pt}{1pt}}
\put(1238,215){\rule{1pt}{1pt}}
\put(620,269){\rule{1pt}{1pt}}
\put(1114,572){\rule{1pt}{1pt}}
\put(595,506){\rule{1pt}{1pt}}
\put(889,759){\rule{1pt}{1pt}}
\put(1101,622){\rule{1pt}{1pt}}
\put(1010,777){\rule{1pt}{1pt}}
\put(1182,363){\rule{1pt}{1pt}}
\put(496,381){\rule{1pt}{1pt}}
\put(660,288){\rule{1pt}{1pt}}
\put(1250,185){\rule{1pt}{1pt}}
\put(918,655){\rule{1pt}{1pt}}
\put(1263,154){\rule{1pt}{1pt}}
\put(1235,222){\rule{1pt}{1pt}}
\put(1045,635){\rule{1pt}{1pt}}
\put(1139,509){\rule{1pt}{1pt}}
\put(897,548){\rule{1pt}{1pt}}
\put(829,685){\rule{1pt}{1pt}}
\put(924,743){\rule{1pt}{1pt}}
\put(285,123){\rule{1pt}{1pt}}
\put(756,661){\rule{1pt}{1pt}}
\put(1195,321){\rule{1pt}{1pt}}
\put(437,211){\rule{1pt}{1pt}}
\put(1100,624){\rule{1pt}{1pt}}
\put(894,658){\rule{1pt}{1pt}}
\put(1009,766){\rule{1pt}{1pt}}
\put(1106,608){\rule{1pt}{1pt}}
\put(680,578){\rule{1pt}{1pt}}
\put(918,658){\rule{1pt}{1pt}}
\put(524,84){\rule{1pt}{1pt}}
\put(738,679){\rule{1pt}{1pt}}
\put(762,717){\rule{1pt}{1pt}}
\put(283,122){\rule{1pt}{1pt}}
\put(955,817){\rule{1pt}{1pt}}
\put(1044,608){\rule{1pt}{1pt}}
\put(704,580){\rule{1pt}{1pt}}
\put(1175,386){\rule{1pt}{1pt}}
\put(755,603){\rule{1pt}{1pt}}
\put(429,99){\rule{1pt}{1pt}}
\put(1182,363){\rule{1pt}{1pt}}
\put(1175,381){\rule{1pt}{1pt}}
\put(784,554){\rule{1pt}{1pt}}
\put(1143,493){\rule{1pt}{1pt}}
\put(660,288){\rule{1pt}{1pt}}
\put(1015,752){\rule{1pt}{1pt}}
\put(520,238){\rule{1pt}{1pt}}
\put(1149,398){\rule{1pt}{1pt}}
\put(931,785){\rule{1pt}{1pt}}
\put(425,126){\rule{1pt}{1pt}}
\put(819,649){\rule{1pt}{1pt}}
\put(756,635){\rule{1pt}{1pt}}
\put(541,442){\rule{1pt}{1pt}}
\put(726,463){\rule{1pt}{1pt}}
\put(619,279){\rule{1pt}{1pt}}
\put(897,562){\rule{1pt}{1pt}}
\put(1001,788){\rule{1pt}{1pt}}
\put(1287,103){\rule{1pt}{1pt}}
\put(748,594){\rule{1pt}{1pt}}
\put(722,622){\rule{1pt}{1pt}}
\put(1257,168){\rule{1pt}{1pt}}
\put(929,755){\rule{1pt}{1pt}}
\put(553,471){\rule{1pt}{1pt}}
\put(985,445){\rule{1pt}{1pt}}
\put(318,104){\rule{1pt}{1pt}}
\put(1206,290){\rule{1pt}{1pt}}
\put(677,519){\rule{1pt}{1pt}}
\put(681,572){\rule{1pt}{1pt}}
\put(530,222){\rule{1pt}{1pt}}
\put(456,152){\rule{1pt}{1pt}}
\put(523,122){\rule{1pt}{1pt}}
\put(959,752){\rule{1pt}{1pt}}
\put(1074,731){\rule{1pt}{1pt}}
\put(972,725){\rule{1pt}{1pt}}
\put(1024,775){\rule{1pt}{1pt}}
\put(1160,439){\rule{1pt}{1pt}}
\put(1217,270){\rule{1pt}{1pt}}
\put(1073,735){\rule{1pt}{1pt}}
\put(1245,197){\rule{1pt}{1pt}}
\put(1034,608){\rule{1pt}{1pt}}
\put(431,159){\rule{1pt}{1pt}}
\put(968,777){\rule{1pt}{1pt}}
\put(895,630){\rule{1pt}{1pt}}
\put(1160,440){\rule{1pt}{1pt}}
\put(769,703){\rule{1pt}{1pt}}
\put(1293,91){\rule{1pt}{1pt}}
\put(944,777){\rule{1pt}{1pt}}
\put(427,135){\rule{1pt}{1pt}}
\put(620,271){\rule{1pt}{1pt}}
\put(1122,542){\rule{1pt}{1pt}}
\put(1014,737){\rule{1pt}{1pt}}
\put(523,124){\rule{1pt}{1pt}}
\put(1083,394){\rule{1pt}{1pt}}
\put(1213,283){\rule{1pt}{1pt}}
\put(846,770){\rule{1pt}{1pt}}
\put(915,559){\rule{1pt}{1pt}}
\put(1225,250){\rule{1pt}{1pt}}
\put(217,95){\rule{1pt}{1pt}}
\put(1093,659){\rule{1pt}{1pt}}
\put(942,710){\rule{1pt}{1pt}}
\put(945,735){\rule{1pt}{1pt}}
\put(742,697){\rule{1pt}{1pt}}
\put(807,334){\rule{1pt}{1pt}}
\put(1254,171){\rule{1pt}{1pt}}
\put(265,107){\rule{1pt}{1pt}}
\put(497,342){\rule{1pt}{1pt}}
\put(592,457){\rule{1pt}{1pt}}
\put(670,434){\rule{1pt}{1pt}}
\put(484,308){\rule{1pt}{1pt}}
\put(660,294){\rule{1pt}{1pt}}
\put(456,135){\rule{1pt}{1pt}}
\put(815,542){\rule{1pt}{1pt}}
\put(558,473){\rule{1pt}{1pt}}
\put(517,318){\rule{1pt}{1pt}}
\put(284,120){\rule{1pt}{1pt}}
\put(248,85){\rule{1pt}{1pt}}
\put(477,236){\rule{1pt}{1pt}}
\put(523,129){\rule{1pt}{1pt}}
\put(452,273){\rule{1pt}{1pt}}
\put(794,257){\rule{1pt}{1pt}}
\put(1152,426){\rule{1pt}{1pt}}
\put(747,707){\rule{1pt}{1pt}}
\put(921,729){\rule{1pt}{1pt}}
\put(926,763){\rule{1pt}{1pt}}
\put(659,277){\rule{1pt}{1pt}}
\put(894,655){\rule{1pt}{1pt}}
\put(701,356){\rule{1pt}{1pt}}
\put(949,742){\rule{1pt}{1pt}}
\put(939,783){\rule{1pt}{1pt}}
\put(935,756){\rule{1pt}{1pt}}
\put(1100,624){\rule{1pt}{1pt}}
\put(809,387){\rule{1pt}{1pt}}
\put(768,699){\rule{1pt}{1pt}}
\put(548,469){\rule{1pt}{1pt}}
\put(783,576){\rule{1pt}{1pt}}
\put(1064,713){\rule{1pt}{1pt}}
\put(997,748){\rule{1pt}{1pt}}
\put(446,262){\rule{1pt}{1pt}}
\put(924,701){\rule{1pt}{1pt}}
\put(927,759){\rule{1pt}{1pt}}
\put(439,189){\rule{1pt}{1pt}}
\put(822,702){\rule{1pt}{1pt}}
\put(825,655){\rule{1pt}{1pt}}
\put(1034,601){\rule{1pt}{1pt}}
\put(315,109){\rule{1pt}{1pt}}
\put(1170,394){\rule{1pt}{1pt}}
\put(451,266){\rule{1pt}{1pt}}
\put(1025,776){\rule{1pt}{1pt}}
\put(723,598){\rule{1pt}{1pt}}
\put(660,294){\rule{1pt}{1pt}}
\put(1176,380){\rule{1pt}{1pt}}
\put(612,402){\rule{1pt}{1pt}}
\put(922,743){\rule{1pt}{1pt}}
\put(1160,439){\rule{1pt}{1pt}}
\put(1019,790){\rule{1pt}{1pt}}
\put(994,748){\rule{1pt}{1pt}}
\put(963,802){\rule{1pt}{1pt}}
\put(823,707){\rule{1pt}{1pt}}
\put(593,488){\rule{1pt}{1pt}}
\put(935,745){\rule{1pt}{1pt}}
\put(1207,285){\rule{1pt}{1pt}}
\put(1149,399){\rule{1pt}{1pt}}
\put(915,559){\rule{1pt}{1pt}}
\put(967,784){\rule{1pt}{1pt}}
\put(1293,90){\rule{1pt}{1pt}}
\put(312,93){\rule{1pt}{1pt}}
\put(685,546){\rule{1pt}{1pt}}
\put(1116,562){\rule{1pt}{1pt}}
\put(842,786){\rule{1pt}{1pt}}
\put(897,558){\rule{1pt}{1pt}}
\put(1233,225){\rule{1pt}{1pt}}
\put(560,477){\rule{1pt}{1pt}}
\put(1136,519){\rule{1pt}{1pt}}
\put(915,567){\rule{1pt}{1pt}}
\put(280,99){\rule{1pt}{1pt}}
\put(823,705){\rule{1pt}{1pt}}
\put(579,502){\rule{1pt}{1pt}}
\put(494,369){\rule{1pt}{1pt}}
\put(612,394){\rule{1pt}{1pt}}
\put(1290,98){\rule{1pt}{1pt}}
\put(1232,230){\rule{1pt}{1pt}}
\put(1277,124){\rule{1pt}{1pt}}
\put(450,265){\rule{1pt}{1pt}}
\put(897,557){\rule{1pt}{1pt}}
\put(984,335){\rule{1pt}{1pt}}
\put(1091,662){\rule{1pt}{1pt}}
\put(830,596){\rule{1pt}{1pt}}
\put(927,652){\rule{1pt}{1pt}}
\put(946,664){\rule{1pt}{1pt}}
\put(286,122){\rule{1pt}{1pt}}
\put(1108,596){\rule{1pt}{1pt}}
\put(1063,551){\rule{1pt}{1pt}}
\put(268,124){\rule{1pt}{1pt}}
\put(1212,285){\rule{1pt}{1pt}}
\put(731,459){\rule{1pt}{1pt}}
\put(1241,208){\rule{1pt}{1pt}}
\put(1259,162){\rule{1pt}{1pt}}
\put(660,289){\rule{1pt}{1pt}}
\put(1135,523){\rule{1pt}{1pt}}
\put(1025,774){\rule{1pt}{1pt}}
\put(532,272){\rule{1pt}{1pt}}
\put(1148,398){\rule{1pt}{1pt}}
\put(830,583){\rule{1pt}{1pt}}
\put(506,354){\rule{1pt}{1pt}}
\put(1069,748){\rule{1pt}{1pt}}
\put(737,674){\rule{1pt}{1pt}}
\put(745,706){\rule{1pt}{1pt}}
\put(1296,86){\rule{1pt}{1pt}}
\put(803,167){\rule{1pt}{1pt}}
\put(1242,202){\rule{1pt}{1pt}}
\put(1064,657){\rule{1pt}{1pt}}
\put(1130,538){\rule{1pt}{1pt}}
\put(272,111){\rule{1pt}{1pt}}
\put(977,559){\rule{1pt}{1pt}}
\put(512,390){\rule{1pt}{1pt}}
\put(435,192){\rule{1pt}{1pt}}
\put(1133,527){\rule{1pt}{1pt}}
\put(984,388){\rule{1pt}{1pt}}
\put(1113,575){\rule{1pt}{1pt}}
\put(783,560){\rule{1pt}{1pt}}
\put(978,513){\rule{1pt}{1pt}}
\put(550,404){\rule{1pt}{1pt}}
\put(608,433){\rule{1pt}{1pt}}
\put(287,121){\rule{1pt}{1pt}}
\put(984,341){\rule{1pt}{1pt}}
\put(485,314){\rule{1pt}{1pt}}
\put(766,525){\rule{1pt}{1pt}}
\put(619,286){\rule{1pt}{1pt}}
\put(1204,297){\rule{1pt}{1pt}}
\put(539,395){\rule{1pt}{1pt}}
\put(758,708){\rule{1pt}{1pt}}
\put(878,792){\rule{1pt}{1pt}}
\put(861,710){\rule{1pt}{1pt}}
\put(477,241){\rule{1pt}{1pt}}
\put(1022,701){\rule{1pt}{1pt}}
\put(738,694){\rule{1pt}{1pt}}
\put(731,415){\rule{1pt}{1pt}}
\put(882,768){\rule{1pt}{1pt}}
\put(617,317){\rule{1pt}{1pt}}
\put(678,507){\rule{1pt}{1pt}}
\put(1085,444){\rule{1pt}{1pt}}
\put(592,456){\rule{1pt}{1pt}}
\put(897,566){\rule{1pt}{1pt}}
\put(763,713){\rule{1pt}{1pt}}
\put(864,788){\rule{1pt}{1pt}}
\put(999,787){\rule{1pt}{1pt}}
\put(727,297){\rule{1pt}{1pt}}
\put(1168,401){\rule{1pt}{1pt}}
\put(557,351){\rule{1pt}{1pt}}
\put(758,701){\rule{1pt}{1pt}}
\put(883,644){\rule{1pt}{1pt}}
\put(1016,791){\rule{1pt}{1pt}}
\put(732,501){\rule{1pt}{1pt}}
\put(752,529){\rule{1pt}{1pt}}
\put(1058,768){\rule{1pt}{1pt}}
\put(518,288){\rule{1pt}{1pt}}
\put(1225,248){\rule{1pt}{1pt}}
\put(759,713){\rule{1pt}{1pt}}
\put(256,116){\rule{1pt}{1pt}}
\put(826,730){\rule{1pt}{1pt}}
\put(825,654){\rule{1pt}{1pt}}
\put(1202,301){\rule{1pt}{1pt}}
\put(889,759){\rule{1pt}{1pt}}
\put(668,414){\rule{1pt}{1pt}}
\put(250,85){\rule{1pt}{1pt}}
\put(1123,560){\rule{1pt}{1pt}}
\put(916,587){\rule{1pt}{1pt}}
\put(789,405){\rule{1pt}{1pt}}
\put(789,404){\rule{1pt}{1pt}}
\put(605,402){\rule{1pt}{1pt}}
\put(693,620){\rule{1pt}{1pt}}
\put(438,212){\rule{1pt}{1pt}}
\put(816,568){\rule{1pt}{1pt}}
\put(686,469){\rule{1pt}{1pt}}
\put(292,125){\rule{1pt}{1pt}}
\put(566,201){\rule{1pt}{1pt}}
\put(1065,764){\rule{1pt}{1pt}}
\put(866,791){\rule{1pt}{1pt}}
\put(783,582){\rule{1pt}{1pt}}
\put(986,483){\rule{1pt}{1pt}}
\put(694,628){\rule{1pt}{1pt}}
\put(884,740){\rule{1pt}{1pt}}
\put(727,346){\rule{1pt}{1pt}}
\put(704,604){\rule{1pt}{1pt}}
\put(284,114){\rule{1pt}{1pt}}
\put(510,392){\rule{1pt}{1pt}}
\put(961,799){\rule{1pt}{1pt}}
\put(258,120){\rule{1pt}{1pt}}
\put(618,304){\rule{1pt}{1pt}}
\put(620,265){\rule{1pt}{1pt}}
\put(599,482){\rule{1pt}{1pt}}
\put(690,593){\rule{1pt}{1pt}}
\put(1197,318){\rule{1pt}{1pt}}
\put(788,433){\rule{1pt}{1pt}}
\put(743,686){\rule{1pt}{1pt}}
\put(723,559){\rule{1pt}{1pt}}
\put(957,813){\rule{1pt}{1pt}}
\put(984,350){\rule{1pt}{1pt}}
\put(671,392){\rule{1pt}{1pt}}
\put(1136,518){\rule{1pt}{1pt}}
\put(1029,659){\rule{1pt}{1pt}}
\put(1133,532){\rule{1pt}{1pt}}
\put(1056,753){\rule{1pt}{1pt}}
\put(227,87){\rule{1pt}{1pt}}
\put(1257,167){\rule{1pt}{1pt}}
\put(527,85){\rule{1pt}{1pt}}
\put(828,734){\rule{1pt}{1pt}}
\put(317,86){\rule{1pt}{1pt}}
\put(317,106){\rule{1pt}{1pt}}
\put(1066,738){\rule{1pt}{1pt}}
\put(695,632){\rule{1pt}{1pt}}
\put(796,166){\rule{1pt}{1pt}}
\put(496,384){\rule{1pt}{1pt}}
\put(1143,487){\rule{1pt}{1pt}}
\put(734,614){\rule{1pt}{1pt}}
\put(1181,367){\rule{1pt}{1pt}}
\put(1249,187){\rule{1pt}{1pt}}
\put(425,125){\rule{1pt}{1pt}}
\put(1209,282){\rule{1pt}{1pt}}
\put(554,483){\rule{1pt}{1pt}}
\put(270,104){\rule{1pt}{1pt}}
\put(493,389){\rule{1pt}{1pt}}
\put(214,89){\rule{1pt}{1pt}}
\put(1231,232){\rule{1pt}{1pt}}
\put(730,359){\rule{1pt}{1pt}}
\put(719,623){\rule{1pt}{1pt}}
\put(539,411){\rule{1pt}{1pt}}
\put(929,778){\rule{1pt}{1pt}}
\put(1165,415){\rule{1pt}{1pt}}
\put(614,357){\rule{1pt}{1pt}}
\put(1229,238){\rule{1pt}{1pt}}
\put(1004,782){\rule{1pt}{1pt}}
\put(809,381){\rule{1pt}{1pt}}
\put(487,395){\rule{1pt}{1pt}}
\put(694,613){\rule{1pt}{1pt}}
\put(1145,484){\rule{1pt}{1pt}}
\put(773,620){\rule{1pt}{1pt}}
\put(427,108){\rule{1pt}{1pt}}
\put(1059,739){\rule{1pt}{1pt}}
\put(1203,299){\rule{1pt}{1pt}}
\put(1175,386){\rule{1pt}{1pt}}
\put(915,558){\rule{1pt}{1pt}}
\put(421,94){\rule{1pt}{1pt}}
\put(566,265){\rule{1pt}{1pt}}
\put(862,747){\rule{1pt}{1pt}}
\put(752,635){\rule{1pt}{1pt}}
\put(974,685){\rule{1pt}{1pt}}
\put(423,110){\rule{1pt}{1pt}}
\put(1140,506){\rule{1pt}{1pt}}
\put(988,589){\rule{1pt}{1pt}}
\put(1228,242){\rule{1pt}{1pt}}
\put(1220,263){\rule{1pt}{1pt}}
\put(942,709){\rule{1pt}{1pt}}
\put(975,620){\rule{1pt}{1pt}}
\put(290,120){\rule{1pt}{1pt}}
\put(1134,529){\rule{1pt}{1pt}}
\put(1144,485){\rule{1pt}{1pt}}
\put(291,113){\rule{1pt}{1pt}}
\put(1016,783){\rule{1pt}{1pt}}
\put(301,83){\rule{1pt}{1pt}}
\put(572,474){\rule{1pt}{1pt}}
\put(1167,405){\rule{1pt}{1pt}}
\put(620,271){\rule{1pt}{1pt}}
\put(1178,371){\rule{1pt}{1pt}}
\put(1191,336){\rule{1pt}{1pt}}
\put(502,396){\rule{1pt}{1pt}}
\put(959,751){\rule{1pt}{1pt}}
\put(465,288){\rule{1pt}{1pt}}
\put(873,744){\rule{1pt}{1pt}}
\put(580,503){\rule{1pt}{1pt}}
\put(1069,740){\rule{1pt}{1pt}}
\put(1267,146){\rule{1pt}{1pt}}
\put(1294,89){\rule{1pt}{1pt}}
\put(819,641){\rule{1pt}{1pt}}
\put(1034,613){\rule{1pt}{1pt}}
\put(952,805){\rule{1pt}{1pt}}
\put(451,264){\rule{1pt}{1pt}}
\put(1149,397){\rule{1pt}{1pt}}
\put(527,88){\rule{1pt}{1pt}}
\put(1044,607){\rule{1pt}{1pt}}
\put(1234,226){\rule{1pt}{1pt}}
\put(543,429){\rule{1pt}{1pt}}
\put(992,720){\rule{1pt}{1pt}}
\put(959,758){\rule{1pt}{1pt}}
\put(269,92){\rule{1pt}{1pt}}
\put(1076,720){\rule{1pt}{1pt}}
\put(1206,291){\rule{1pt}{1pt}}
\put(1135,524){\rule{1pt}{1pt}}
\put(1153,459){\rule{1pt}{1pt}}
\put(1296,87){\rule{1pt}{1pt}}
\put(459,297){\rule{1pt}{1pt}}
\put(700,475){\rule{1pt}{1pt}}
\put(594,484){\rule{1pt}{1pt}}
\put(748,644){\rule{1pt}{1pt}}
\put(967,789){\rule{1pt}{1pt}}
\put(1109,597){\rule{1pt}{1pt}}
\put(546,465){\rule{1pt}{1pt}}
\put(684,569){\rule{1pt}{1pt}}
\put(499,208){\rule{1pt}{1pt}}
\put(984,350){\rule{1pt}{1pt}}
\put(751,677){\rule{1pt}{1pt}}
\put(1024,770){\rule{1pt}{1pt}}
\put(705,645){\rule{1pt}{1pt}}
\put(734,620){\rule{1pt}{1pt}}
\put(1060,763){\rule{1pt}{1pt}}
\put(198,84){\rule{1pt}{1pt}}
\put(305,117){\rule{1pt}{1pt}}
\put(524,84){\rule{1pt}{1pt}}
\put(897,546){\rule{1pt}{1pt}}
\put(265,117){\rule{1pt}{1pt}}
\put(1079,642){\rule{1pt}{1pt}}
\put(472,334){\rule{1pt}{1pt}}
\put(462,296){\rule{1pt}{1pt}}
\put(969,733){\rule{1pt}{1pt}}
\put(1016,776){\rule{1pt}{1pt}}
\put(1192,331){\rule{1pt}{1pt}}
\put(971,751){\rule{1pt}{1pt}}
\put(421,94){\rule{1pt}{1pt}}
\put(533,320){\rule{1pt}{1pt}}
\put(1142,497){\rule{1pt}{1pt}}
\put(455,113){\rule{1pt}{1pt}}
\put(984,374){\rule{1pt}{1pt}}
\put(669,434){\rule{1pt}{1pt}}
\put(1155,453){\rule{1pt}{1pt}}
\put(603,459){\rule{1pt}{1pt}}
\put(1151,465){\rule{1pt}{1pt}}
\put(741,703){\rule{1pt}{1pt}}
\put(424,113){\rule{1pt}{1pt}}
\put(270,119){\rule{1pt}{1pt}}
\put(475,271){\rule{1pt}{1pt}}
\put(974,668){\rule{1pt}{1pt}}
\put(496,388){\rule{1pt}{1pt}}
\put(494,387){\rule{1pt}{1pt}}
\put(919,675){\rule{1pt}{1pt}}
\put(317,105){\rule{1pt}{1pt}}
\put(1106,608){\rule{1pt}{1pt}}
\put(796,171){\rule{1pt}{1pt}}
\put(1138,511){\rule{1pt}{1pt}}
\put(620,271){\rule{1pt}{1pt}}
\put(1020,791){\rule{1pt}{1pt}}
\put(284,124){\rule{1pt}{1pt}}
\put(1008,784){\rule{1pt}{1pt}}
\put(1077,693){\rule{1pt}{1pt}}
\put(1288,100){\rule{1pt}{1pt}}
\put(884,776){\rule{1pt}{1pt}}
\put(776,672){\rule{1pt}{1pt}}
\put(787,460){\rule{1pt}{1pt}}
\put(620,268){\rule{1pt}{1pt}}
\put(1016,775){\rule{1pt}{1pt}}
\put(949,682){\rule{1pt}{1pt}}
\put(829,629){\rule{1pt}{1pt}}
\put(554,485){\rule{1pt}{1pt}}
\put(897,564){\rule{1pt}{1pt}}
\put(1030,741){\rule{1pt}{1pt}}
\put(756,637){\rule{1pt}{1pt}}
\put(732,502){\rule{1pt}{1pt}}
\put(283,123){\rule{1pt}{1pt}}
\put(1075,728){\rule{1pt}{1pt}}
\put(727,295){\rule{1pt}{1pt}}
\put(1224,252){\rule{1pt}{1pt}}
\put(683,589){\rule{1pt}{1pt}}
\put(1114,570){\rule{1pt}{1pt}}
\put(1091,657){\rule{1pt}{1pt}}
\put(1190,336){\rule{1pt}{1pt}}
\put(568,263){\rule{1pt}{1pt}}
\put(882,757){\rule{1pt}{1pt}}
\put(992,706){\rule{1pt}{1pt}}
\put(573,481){\rule{1pt}{1pt}}
\put(223,93){\rule{1pt}{1pt}}
\put(1062,625){\rule{1pt}{1pt}}
\put(1228,240){\rule{1pt}{1pt}}
\put(1257,167){\rule{1pt}{1pt}}
\put(292,122){\rule{1pt}{1pt}}
\put(543,449){\rule{1pt}{1pt}}
\put(284,104){\rule{1pt}{1pt}}
\put(258,115){\rule{1pt}{1pt}}
\put(666,392){\rule{1pt}{1pt}}
\put(620,266){\rule{1pt}{1pt}}
\put(1044,605){\rule{1pt}{1pt}}
\put(558,473){\rule{1pt}{1pt}}
\put(488,399){\rule{1pt}{1pt}}
\put(1122,557){\rule{1pt}{1pt}}
\put(458,255){\rule{1pt}{1pt}}
\put(582,496){\rule{1pt}{1pt}}
\put(569,332){\rule{1pt}{1pt}}
\put(732,495){\rule{1pt}{1pt}}
\put(959,754){\rule{1pt}{1pt}}
\put(1044,609){\rule{1pt}{1pt}}
\put(1124,557){\rule{1pt}{1pt}}
\put(1268,143){\rule{1pt}{1pt}}
\put(1221,262){\rule{1pt}{1pt}}
\put(790,395){\rule{1pt}{1pt}}
\put(922,742){\rule{1pt}{1pt}}
\put(863,777){\rule{1pt}{1pt}}
\put(433,176){\rule{1pt}{1pt}}
\put(1285,107){\rule{1pt}{1pt}}
\put(549,458){\rule{1pt}{1pt}}
\put(735,609){\rule{1pt}{1pt}}
\put(1288,100){\rule{1pt}{1pt}}
\put(1088,595){\rule{1pt}{1pt}}
\put(758,680){\rule{1pt}{1pt}}
\put(1114,572){\rule{1pt}{1pt}}
\put(897,562){\rule{1pt}{1pt}}
\put(1272,135){\rule{1pt}{1pt}}
\put(794,263){\rule{1pt}{1pt}}
\put(1274,129){\rule{1pt}{1pt}}
\put(706,652){\rule{1pt}{1pt}}
\put(250,100){\rule{1pt}{1pt}}
\put(1122,559){\rule{1pt}{1pt}}
\put(811,431){\rule{1pt}{1pt}}
\put(772,624){\rule{1pt}{1pt}}
\put(476,256){\rule{1pt}{1pt}}
\put(1166,411){\rule{1pt}{1pt}}
\put(726,432){\rule{1pt}{1pt}}
\put(1091,662){\rule{1pt}{1pt}}
\put(1143,494){\rule{1pt}{1pt}}
\put(922,742){\rule{1pt}{1pt}}
\put(448,278){\rule{1pt}{1pt}}
\put(752,635){\rule{1pt}{1pt}}
\put(300,113){\rule{1pt}{1pt}}
\put(766,486){\rule{1pt}{1pt}}
\put(556,483){\rule{1pt}{1pt}}
\put(830,581){\rule{1pt}{1pt}}
\put(1130,537){\rule{1pt}{1pt}}
\put(977,545){\rule{1pt}{1pt}}
\put(971,735){\rule{1pt}{1pt}}
\put(620,273){\rule{1pt}{1pt}}
\put(659,282){\rule{1pt}{1pt}}
\put(920,707){\rule{1pt}{1pt}}
\put(300,101){\rule{1pt}{1pt}}
\put(1095,638){\rule{1pt}{1pt}}
\put(1154,455){\rule{1pt}{1pt}}
\put(262,96){\rule{1pt}{1pt}}
\put(891,722){\rule{1pt}{1pt}}
\put(1011,716){\rule{1pt}{1pt}}
\put(289,114){\rule{1pt}{1pt}}
\put(1054,760){\rule{1pt}{1pt}}
\put(469,335){\rule{1pt}{1pt}}
\put(766,484){\rule{1pt}{1pt}}
\put(768,700){\rule{1pt}{1pt}}
\put(757,693){\rule{1pt}{1pt}}
\put(780,610){\rule{1pt}{1pt}}
\put(696,633){\rule{1pt}{1pt}}
\put(1258,165){\rule{1pt}{1pt}}
\put(516,363){\rule{1pt}{1pt}}
\put(815,548){\rule{1pt}{1pt}}
\put(534,326){\rule{1pt}{1pt}}
\put(704,625){\rule{1pt}{1pt}}
\put(787,459){\rule{1pt}{1pt}}
\put(887,742){\rule{1pt}{1pt}}
\put(995,749){\rule{1pt}{1pt}}
\put(808,354){\rule{1pt}{1pt}}
\put(553,473){\rule{1pt}{1pt}}
\put(1183,355){\rule{1pt}{1pt}}
\put(735,664){\rule{1pt}{1pt}}
\put(1170,395){\rule{1pt}{1pt}}
\put(269,108){\rule{1pt}{1pt}}
\put(523,136){\rule{1pt}{1pt}}
\put(781,616){\rule{1pt}{1pt}}
\put(1044,607){\rule{1pt}{1pt}}
\put(1166,415){\rule{1pt}{1pt}}
\put(1085,434){\rule{1pt}{1pt}}
\put(215,94){\rule{1pt}{1pt}}
\put(915,153){\rule{1pt}{1pt}}
\put(511,362){\rule{1pt}{1pt}}
\put(1175,386){\rule{1pt}{1pt}}
\put(670,444){\rule{1pt}{1pt}}
\put(1266,148){\rule{1pt}{1pt}}
\put(1138,516){\rule{1pt}{1pt}}
\put(988,561){\rule{1pt}{1pt}}
\put(1079,646){\rule{1pt}{1pt}}
\put(788,441){\rule{1pt}{1pt}}
\put(660,279){\rule{1pt}{1pt}}
\put(1166,416){\rule{1pt}{1pt}}
\put(266,103){\rule{1pt}{1pt}}
\put(948,648){\rule{1pt}{1pt}}
\put(1218,262){\rule{1pt}{1pt}}
\put(544,336){\rule{1pt}{1pt}}
\put(925,762){\rule{1pt}{1pt}}
\put(256,115){\rule{1pt}{1pt}}
\put(1050,734){\rule{1pt}{1pt}}
\put(720,662){\rule{1pt}{1pt}}
\put(954,805){\rule{1pt}{1pt}}
\put(295,88){\rule{1pt}{1pt}}
\put(589,488){\rule{1pt}{1pt}}
\put(841,734){\rule{1pt}{1pt}}
\put(789,426){\rule{1pt}{1pt}}
\put(772,635){\rule{1pt}{1pt}}
\put(946,636){\rule{1pt}{1pt}}
\put(803,166){\rule{1pt}{1pt}}
\put(661,299){\rule{1pt}{1pt}}
\put(1116,560){\rule{1pt}{1pt}}
\put(1275,127){\rule{1pt}{1pt}}
\put(1134,526){\rule{1pt}{1pt}}
\put(1177,378){\rule{1pt}{1pt}}
\put(659,283){\rule{1pt}{1pt}}
\put(542,443){\rule{1pt}{1pt}}
\put(620,269){\rule{1pt}{1pt}}
\put(865,790){\rule{1pt}{1pt}}
\put(1106,606){\rule{1pt}{1pt}}
\put(254,112){\rule{1pt}{1pt}}
\put(834,713){\rule{1pt}{1pt}}
\put(918,658){\rule{1pt}{1pt}}
\put(919,685){\rule{1pt}{1pt}}
\put(667,407){\rule{1pt}{1pt}}
\put(808,368){\rule{1pt}{1pt}}
\put(1223,256){\rule{1pt}{1pt}}
\put(1140,506){\rule{1pt}{1pt}}
\put(263,122){\rule{1pt}{1pt}}
\put(1222,254){\rule{1pt}{1pt}}
\put(619,281){\rule{1pt}{1pt}}
\put(1044,602){\rule{1pt}{1pt}}
\put(1175,387){\rule{1pt}{1pt}}
\put(282,115){\rule{1pt}{1pt}}
\put(1088,601){\rule{1pt}{1pt}}
\put(1287,102){\rule{1pt}{1pt}}
\put(1116,562){\rule{1pt}{1pt}}
\put(1104,613){\rule{1pt}{1pt}}
\put(477,241){\rule{1pt}{1pt}}
\put(1191,337){\rule{1pt}{1pt}}
\put(771,693){\rule{1pt}{1pt}}
\put(1067,670){\rule{1pt}{1pt}}
\put(932,780){\rule{1pt}{1pt}}
\put(682,583){\rule{1pt}{1pt}}
\put(692,624){\rule{1pt}{1pt}}
\put(514,375){\rule{1pt}{1pt}}
\put(890,750){\rule{1pt}{1pt}}
\put(1089,641){\rule{1pt}{1pt}}
\put(897,562){\rule{1pt}{1pt}}
\put(752,632){\rule{1pt}{1pt}}
\put(714,667){\rule{1pt}{1pt}}
\put(763,681){\rule{1pt}{1pt}}
\put(917,627){\rule{1pt}{1pt}}
\put(759,716){\rule{1pt}{1pt}}
\put(1136,522){\rule{1pt}{1pt}}
\put(524,84){\rule{1pt}{1pt}}
\put(1085,394){\rule{1pt}{1pt}}
\put(303,114){\rule{1pt}{1pt}}
\put(711,668){\rule{1pt}{1pt}}
\put(740,657){\rule{1pt}{1pt}}
\put(662,325){\rule{1pt}{1pt}}
\put(1089,656){\rule{1pt}{1pt}}
\put(1096,647){\rule{1pt}{1pt}}
\put(813,504){\rule{1pt}{1pt}}
\put(1238,215){\rule{1pt}{1pt}}
\put(464,318){\rule{1pt}{1pt}}
\put(662,324){\rule{1pt}{1pt}}
\put(888,669){\rule{1pt}{1pt}}
\put(681,584){\rule{1pt}{1pt}}
\put(1056,739){\rule{1pt}{1pt}}
\put(495,392){\rule{1pt}{1pt}}
\put(428,145){\rule{1pt}{1pt}}
\put(253,105){\rule{1pt}{1pt}}
\put(951,793){\rule{1pt}{1pt}}
\put(1094,651){\rule{1pt}{1pt}}
\put(544,426){\rule{1pt}{1pt}}
\put(287,125){\rule{1pt}{1pt}}
\put(1195,325){\rule{1pt}{1pt}}
\put(1196,319){\rule{1pt}{1pt}}
\put(1005,793){\rule{1pt}{1pt}}
\put(290,128){\rule{1pt}{1pt}}
\put(1013,594){\rule{1pt}{1pt}}
\put(958,796){\rule{1pt}{1pt}}
\put(684,591){\rule{1pt}{1pt}}
\put(774,646){\rule{1pt}{1pt}}
\put(707,661){\rule{1pt}{1pt}}
\put(590,490){\rule{1pt}{1pt}}
\put(439,191){\rule{1pt}{1pt}}
\put(660,291){\rule{1pt}{1pt}}
\put(711,663){\rule{1pt}{1pt}}
\put(280,86){\rule{1pt}{1pt}}
\put(1095,644){\rule{1pt}{1pt}}
\put(857,436){\rule{1pt}{1pt}}
\put(679,549){\rule{1pt}{1pt}}
\put(949,712){\rule{1pt}{1pt}}
\put(566,200){\rule{1pt}{1pt}}
\put(723,614){\rule{1pt}{1pt}}
\put(709,673){\rule{1pt}{1pt}}
\put(934,587){\rule{1pt}{1pt}}
\put(1226,245){\rule{1pt}{1pt}}
\put(796,180){\rule{1pt}{1pt}}
\put(474,318){\rule{1pt}{1pt}}
\put(813,504){\rule{1pt}{1pt}}
\put(577,501){\rule{1pt}{1pt}}
\put(891,719){\rule{1pt}{1pt}}
\put(968,774){\rule{1pt}{1pt}}
\put(521,212){\rule{1pt}{1pt}}
\put(1202,301){\rule{1pt}{1pt}}
\put(921,663){\rule{1pt}{1pt}}
\put(944,765){\rule{1pt}{1pt}}
\put(925,772){\rule{1pt}{1pt}}
\put(611,416){\rule{1pt}{1pt}}
\put(1283,109){\rule{1pt}{1pt}}
\put(620,275){\rule{1pt}{1pt}}
\put(1209,282){\rule{1pt}{1pt}}
\put(258,118){\rule{1pt}{1pt}}
\put(916,596){\rule{1pt}{1pt}}
\put(915,563){\rule{1pt}{1pt}}
\put(528,125){\rule{1pt}{1pt}}
\put(1208,281){\rule{1pt}{1pt}}
\put(1149,398){\rule{1pt}{1pt}}
\put(1123,561){\rule{1pt}{1pt}}
\put(821,684){\rule{1pt}{1pt}}
\put(1012,563){\rule{1pt}{1pt}}
\put(843,701){\rule{1pt}{1pt}}
\put(264,96){\rule{1pt}{1pt}}
\put(660,290){\rule{1pt}{1pt}}
\put(484,291){\rule{1pt}{1pt}}
\put(514,352){\rule{1pt}{1pt}}
\put(965,787){\rule{1pt}{1pt}}
\put(689,581){\rule{1pt}{1pt}}
\put(490,374){\rule{1pt}{1pt}}
\put(1248,190){\rule{1pt}{1pt}}
\put(452,258){\rule{1pt}{1pt}}
\put(1209,282){\rule{1pt}{1pt}}
\put(663,345){\rule{1pt}{1pt}}
\put(501,386){\rule{1pt}{1pt}}
\put(915,549){\rule{1pt}{1pt}}
\put(830,581){\rule{1pt}{1pt}}
\put(664,365){\rule{1pt}{1pt}}
\put(1059,767){\rule{1pt}{1pt}}
\put(1034,608){\rule{1pt}{1pt}}
\put(1176,380){\rule{1pt}{1pt}}
\put(214,90){\rule{1pt}{1pt}}
\put(1141,503){\rule{1pt}{1pt}}
\put(660,289){\rule{1pt}{1pt}}
\put(759,675){\rule{1pt}{1pt}}
\put(822,704){\rule{1pt}{1pt}}
\put(1103,590){\rule{1pt}{1pt}}
\put(222,93){\rule{1pt}{1pt}}
\put(843,781){\rule{1pt}{1pt}}
\put(897,565){\rule{1pt}{1pt}}
\put(317,95){\rule{1pt}{1pt}}
\put(619,278){\rule{1pt}{1pt}}
\put(270,118){\rule{1pt}{1pt}}
\put(1265,146){\rule{1pt}{1pt}}
\put(673,498){\rule{1pt}{1pt}}
\put(562,459){\rule{1pt}{1pt}}
\put(1134,523){\rule{1pt}{1pt}}
\put(510,354){\rule{1pt}{1pt}}
\put(1177,378){\rule{1pt}{1pt}}
\put(574,471){\rule{1pt}{1pt}}
\put(713,602){\rule{1pt}{1pt}}
\put(773,622){\rule{1pt}{1pt}}
\put(1166,409){\rule{1pt}{1pt}}
\put(1280,118){\rule{1pt}{1pt}}
\put(680,531){\rule{1pt}{1pt}}
\put(230,88){\rule{1pt}{1pt}}
\put(1244,198){\rule{1pt}{1pt}}
\put(1189,335){\rule{1pt}{1pt}}
\put(576,489){\rule{1pt}{1pt}}
\put(810,402){\rule{1pt}{1pt}}
\put(867,772){\rule{1pt}{1pt}}
\put(926,771){\rule{1pt}{1pt}}
\put(325,87){\rule{1pt}{1pt}}
\put(1027,772){\rule{1pt}{1pt}}
\put(761,712){\rule{1pt}{1pt}}
\put(1182,364){\rule{1pt}{1pt}}
\put(806,271){\rule{1pt}{1pt}}
\put(425,124){\rule{1pt}{1pt}}
\put(544,420){\rule{1pt}{1pt}}
\put(1167,408){\rule{1pt}{1pt}}
\put(841,694){\rule{1pt}{1pt}}
\put(1093,661){\rule{1pt}{1pt}}
\put(537,403){\rule{1pt}{1pt}}
\put(1122,554){\rule{1pt}{1pt}}
\put(1116,562){\rule{1pt}{1pt}}
\put(228,95){\rule{1pt}{1pt}}
\put(423,110){\rule{1pt}{1pt}}
\put(1116,562){\rule{1pt}{1pt}}
\put(822,706){\rule{1pt}{1pt}}
\put(1093,659){\rule{1pt}{1pt}}
\put(265,123){\rule{1pt}{1pt}}
\put(1096,646){\rule{1pt}{1pt}}
\put(995,741){\rule{1pt}{1pt}}
\put(765,548){\rule{1pt}{1pt}}
\put(1106,599){\rule{1pt}{1pt}}
\put(848,778){\rule{1pt}{1pt}}
\put(1284,108){\rule{1pt}{1pt}}
\put(1034,603){\rule{1pt}{1pt}}
\put(1212,283){\rule{1pt}{1pt}}
\put(720,655){\rule{1pt}{1pt}}
\put(1187,344){\rule{1pt}{1pt}}
\put(1281,116){\rule{1pt}{1pt}}
\put(1192,332){\rule{1pt}{1pt}}
\put(953,815){\rule{1pt}{1pt}}
\put(1280,117){\rule{1pt}{1pt}}
\put(1002,785){\rule{1pt}{1pt}}
\put(673,483){\rule{1pt}{1pt}}
\put(870,793){\rule{1pt}{1pt}}
\put(1116,562){\rule{1pt}{1pt}}
\put(426,132){\rule{1pt}{1pt}}
\put(1025,779){\rule{1pt}{1pt}}
\put(726,416){\rule{1pt}{1pt}}
\put(1033,641){\rule{1pt}{1pt}}
\put(1168,401){\rule{1pt}{1pt}}
\put(765,658){\rule{1pt}{1pt}}
\put(878,797){\rule{1pt}{1pt}}
\put(487,314){\rule{1pt}{1pt}}
\put(701,360){\rule{1pt}{1pt}}
\put(792,336){\rule{1pt}{1pt}}
\put(1169,403){\rule{1pt}{1pt}}
\put(619,283){\rule{1pt}{1pt}}
\put(897,557){\rule{1pt}{1pt}}
\put(1116,562){\rule{1pt}{1pt}}
\put(543,449){\rule{1pt}{1pt}}
\put(1056,695){\rule{1pt}{1pt}}
\put(1292,92){\rule{1pt}{1pt}}
\put(720,661){\rule{1pt}{1pt}}
\put(823,672){\rule{1pt}{1pt}}
\put(1255,172){\rule{1pt}{1pt}}
\put(294,96){\rule{1pt}{1pt}}
\put(1165,415){\rule{1pt}{1pt}}
\put(227,95){\rule{1pt}{1pt}}
\put(930,791){\rule{1pt}{1pt}}
\put(660,290){\rule{1pt}{1pt}}
\put(504,404){\rule{1pt}{1pt}}
\put(1244,200){\rule{1pt}{1pt}}
\put(1221,261){\rule{1pt}{1pt}}
\put(1285,108){\rule{1pt}{1pt}}
\put(556,477){\rule{1pt}{1pt}}
\put(975,642){\rule{1pt}{1pt}}
\put(1086,537){\rule{1pt}{1pt}}
\put(938,767){\rule{1pt}{1pt}}
\put(1212,284){\rule{1pt}{1pt}}
\put(1098,631){\rule{1pt}{1pt}}
\put(524,84){\rule{1pt}{1pt}}
\put(1166,413){\rule{1pt}{1pt}}
\put(1263,155){\rule{1pt}{1pt}}
\put(660,291){\rule{1pt}{1pt}}
\put(528,129){\rule{1pt}{1pt}}
\put(849,786){\rule{1pt}{1pt}}
\put(1235,217){\rule{1pt}{1pt}}
\put(469,334){\rule{1pt}{1pt}}
\put(1293,91){\rule{1pt}{1pt}}
\put(1273,133){\rule{1pt}{1pt}}
\put(1186,351){\rule{1pt}{1pt}}
\put(1283,110){\rule{1pt}{1pt}}
\put(1146,484){\rule{1pt}{1pt}}
\put(1116,560){\rule{1pt}{1pt}}
\put(574,366){\rule{1pt}{1pt}}
\put(1131,531){\rule{1pt}{1pt}}
\put(790,389){\rule{1pt}{1pt}}
\put(1286,106){\rule{1pt}{1pt}}
\put(718,663){\rule{1pt}{1pt}}
\put(896,582){\rule{1pt}{1pt}}
\put(509,400){\rule{1pt}{1pt}}
\put(446,271){\rule{1pt}{1pt}}
\put(824,655){\rule{1pt}{1pt}}
\put(587,267){\rule{1pt}{1pt}}
\put(981,338){\rule{1pt}{1pt}}
\put(946,573){\rule{1pt}{1pt}}
\put(710,640){\rule{1pt}{1pt}}
\put(268,105){\rule{1pt}{1pt}}
\put(1090,661){\rule{1pt}{1pt}}
\put(265,126){\rule{1pt}{1pt}}
\put(1188,345){\rule{1pt}{1pt}}
\put(1245,197){\rule{1pt}{1pt}}
\put(954,762){\rule{1pt}{1pt}}
\put(477,240){\rule{1pt}{1pt}}
\put(713,603){\rule{1pt}{1pt}}
\put(254,112){\rule{1pt}{1pt}}
\put(429,100){\rule{1pt}{1pt}}
\put(614,332){\rule{1pt}{1pt}}
\put(1209,283){\rule{1pt}{1pt}}
\put(1270,137){\rule{1pt}{1pt}}
\put(1175,381){\rule{1pt}{1pt}}
\put(879,795){\rule{1pt}{1pt}}
\put(1113,571){\rule{1pt}{1pt}}
\put(873,779){\rule{1pt}{1pt}}
\put(895,608){\rule{1pt}{1pt}}
\put(659,285){\rule{1pt}{1pt}}
\put(1141,503){\rule{1pt}{1pt}}
\put(936,803){\rule{1pt}{1pt}}
\put(739,664){\rule{1pt}{1pt}}
\put(1045,615){\rule{1pt}{1pt}}
\put(897,566){\rule{1pt}{1pt}}
\put(944,758){\rule{1pt}{1pt}}
\put(686,514){\rule{1pt}{1pt}}
\put(725,509){\rule{1pt}{1pt}}
\put(1281,116){\rule{1pt}{1pt}}
\put(438,212){\rule{1pt}{1pt}}
\put(588,392){\rule{1pt}{1pt}}
\put(1138,512){\rule{1pt}{1pt}}
\put(1237,215){\rule{1pt}{1pt}}
\put(752,692){\rule{1pt}{1pt}}
\put(620,273){\rule{1pt}{1pt}}
\put(1014,691){\rule{1pt}{1pt}}
\put(748,590){\rule{1pt}{1pt}}
\put(971,762){\rule{1pt}{1pt}}
\put(620,269){\rule{1pt}{1pt}}
\put(584,466){\rule{1pt}{1pt}}
\put(312,89){\rule{1pt}{1pt}}
\put(742,697){\rule{1pt}{1pt}}
\put(620,274){\rule{1pt}{1pt}}
\put(810,403){\rule{1pt}{1pt}}
\put(431,167){\rule{1pt}{1pt}}
\put(1237,215){\rule{1pt}{1pt}}
\put(300,115){\rule{1pt}{1pt}}
\put(759,719){\rule{1pt}{1pt}}
\put(1270,139){\rule{1pt}{1pt}}
\put(1009,766){\rule{1pt}{1pt}}
\put(715,674){\rule{1pt}{1pt}}
\put(874,567){\rule{1pt}{1pt}}
\put(1061,751){\rule{1pt}{1pt}}
\put(739,672){\rule{1pt}{1pt}}
\put(282,118){\rule{1pt}{1pt}}
\put(991,681){\rule{1pt}{1pt}}
\put(617,313){\rule{1pt}{1pt}}
\put(1134,516){\rule{1pt}{1pt}}
\put(1049,674){\rule{1pt}{1pt}}
\put(447,181){\rule{1pt}{1pt}}
\put(1176,383){\rule{1pt}{1pt}}
\put(992,727){\rule{1pt}{1pt}}
\put(1272,135){\rule{1pt}{1pt}}
\put(1028,771){\rule{1pt}{1pt}}
\put(215,94){\rule{1pt}{1pt}}
\put(897,557){\rule{1pt}{1pt}}
\put(803,167){\rule{1pt}{1pt}}
\put(556,461){\rule{1pt}{1pt}}
\put(437,208){\rule{1pt}{1pt}}
\put(441,191){\rule{1pt}{1pt}}
\put(1293,91){\rule{1pt}{1pt}}
\put(281,108){\rule{1pt}{1pt}}
\put(1293,90){\rule{1pt}{1pt}}
\put(698,611){\rule{1pt}{1pt}}
\put(851,767){\rule{1pt}{1pt}}
\put(849,786){\rule{1pt}{1pt}}
\put(492,333){\rule{1pt}{1pt}}
\put(821,686){\rule{1pt}{1pt}}
\put(452,268){\rule{1pt}{1pt}}
\put(459,296){\rule{1pt}{1pt}}
\put(956,816){\rule{1pt}{1pt}}
\put(976,607){\rule{1pt}{1pt}}
\put(1209,281){\rule{1pt}{1pt}}
\put(313,109){\rule{1pt}{1pt}}
\put(846,770){\rule{1pt}{1pt}}
\put(1177,372){\rule{1pt}{1pt}}
\put(1108,590){\rule{1pt}{1pt}}
\put(1030,750){\rule{1pt}{1pt}}
\put(663,335){\rule{1pt}{1pt}}
\put(1044,607){\rule{1pt}{1pt}}
\put(659,277){\rule{1pt}{1pt}}
\put(484,308){\rule{1pt}{1pt}}
\put(594,430){\rule{1pt}{1pt}}
\put(776,674){\rule{1pt}{1pt}}
\put(846,750){\rule{1pt}{1pt}}
\put(587,271){\rule{1pt}{1pt}}
\put(995,736){\rule{1pt}{1pt}}
\put(691,603){\rule{1pt}{1pt}}
\put(891,648){\rule{1pt}{1pt}}
\put(923,724){\rule{1pt}{1pt}}
\put(915,569){\rule{1pt}{1pt}}
\put(997,759){\rule{1pt}{1pt}}
\put(1238,214){\rule{1pt}{1pt}}
\put(532,267){\rule{1pt}{1pt}}
\put(964,793){\rule{1pt}{1pt}}
\put(1085,425){\rule{1pt}{1pt}}
\put(1175,380){\rule{1pt}{1pt}}
\put(522,146){\rule{1pt}{1pt}}
\put(940,778){\rule{1pt}{1pt}}
\put(746,719){\rule{1pt}{1pt}}
\put(605,322){\rule{1pt}{1pt}}
\put(1131,537){\rule{1pt}{1pt}}
\put(1218,270){\rule{1pt}{1pt}}
\put(1181,365){\rule{1pt}{1pt}}
\put(841,704){\rule{1pt}{1pt}}
\put(705,631){\rule{1pt}{1pt}}
\put(555,487){\rule{1pt}{1pt}}
\put(1276,127){\rule{1pt}{1pt}}
\put(1083,398){\rule{1pt}{1pt}}
\put(897,556){\rule{1pt}{1pt}}
\put(500,372){\rule{1pt}{1pt}}
\put(499,208){\rule{1pt}{1pt}}
\put(970,771){\rule{1pt}{1pt}}
\put(1002,725){\rule{1pt}{1pt}}
\put(619,278){\rule{1pt}{1pt}}
\put(830,542){\rule{1pt}{1pt}}
\put(666,394){\rule{1pt}{1pt}}
\put(259,109){\rule{1pt}{1pt}}
\put(918,658){\rule{1pt}{1pt}}
\put(751,701){\rule{1pt}{1pt}}
\put(1031,660){\rule{1pt}{1pt}}
\put(614,356){\rule{1pt}{1pt}}
\put(1030,752){\rule{1pt}{1pt}}
\put(1214,277){\rule{1pt}{1pt}}
\put(1136,521){\rule{1pt}{1pt}}
\put(1104,608){\rule{1pt}{1pt}}
\put(559,484){\rule{1pt}{1pt}}
\put(722,641){\rule{1pt}{1pt}}
\put(1116,558){\rule{1pt}{1pt}}
\put(472,317){\rule{1pt}{1pt}}
\put(269,123){\rule{1pt}{1pt}}
\put(519,284){\rule{1pt}{1pt}}
\put(1187,344){\rule{1pt}{1pt}}
\put(965,744){\rule{1pt}{1pt}}
\put(1252,180){\rule{1pt}{1pt}}
\put(1066,739){\rule{1pt}{1pt}}
\put(672,452){\rule{1pt}{1pt}}
\put(685,523){\rule{1pt}{1pt}}
\put(534,330){\rule{1pt}{1pt}}
\put(1034,601){\rule{1pt}{1pt}}
\put(818,614){\rule{1pt}{1pt}}
\put(825,656){\rule{1pt}{1pt}}
\put(688,521){\rule{1pt}{1pt}}
\put(453,230){\rule{1pt}{1pt}}
\put(283,117){\rule{1pt}{1pt}}
\put(1140,505){\rule{1pt}{1pt}}
\put(1149,406){\rule{1pt}{1pt}}
\end{picture}

\end{figure}
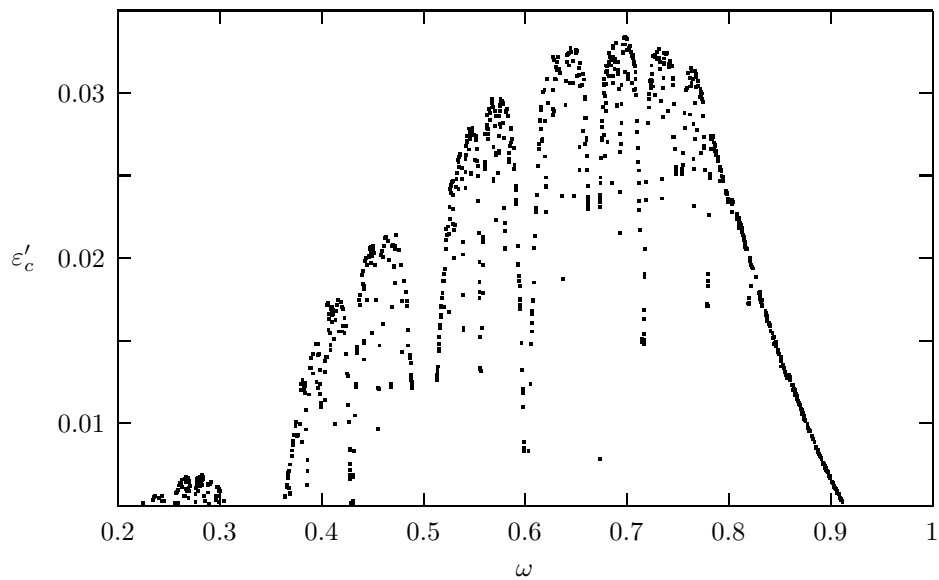

Figure~\ref{fig:epsc} shows the value of $\varepsilon'_c(\omega)$ for
$\omega \in ]0,1[$ determined by the renormalization method with
$L=20$.  A first
remark is that the last invariant torus is not the golden mean one as
it is the case for the standard map and for Escande's paradigm
Hamiltonian~\cite{chan00a}~: The torus with frequency
$\gamma=(\sqrt{5}-1)/2=[1,1,1,\ldots]$ is broken for
$\varepsilon'_c(\gamma)\approx 0.02995$ whereas for instance the torus
with frequency $\omega_1=(\gamma +3)/5=[1,2,1,1,1,\ldots]$ persists until
$\varepsilon'_c(\omega_1)\approx 0.03163$. These
two values coincide up to numerical precision with the ones obtained
by Greene's residue criterion~:
$\varepsilon^{(G)}_c(\gamma)\approx 0.0299$ and
$\varepsilon_c^{(G)}(\omega_1)\approx 0.0316$.\\
The last KAM torus is broken for $\varepsilon^* \approx 0.03334$. Its
frequency is equal to $\omega_2\approx 0.6976$. The critical threshold
of the break-up of this torus with frequency $\omega_2$ obtained by
Greene's residue criterion coincides with $\varepsilon^*$ up to
numerical precision~: $\varepsilon_c^{(G)}(\omega_2)\approx 0.0333$.
For $\varepsilon > \varepsilon^*$, there are no invariant tori
left, and large scale
stochasticity occurs~: trajectories can go from the resonance 1:1
(located at $A_1=1$) to the resonance 0:1 (located at $A_1=0$).\\
Between two neighboring main resonances 1:$n$ and 1:$n$+2,
quasiperiodic motion with frequency $\omega \in ]1/(n+2),1/n[$ can
occur for small $\varepsilon'$. If $\varepsilon'$ is greater than some
value $\varepsilon_n$, there is no longer any quasiperiodic motion in
between these two resonances (and some chaotic trajectories can go
from one resonance to the other). For $n=1$, this value is equal to
$\varepsilon^*$ for which all rotational invariant tori are broken since the
last invariant torus to break-up is located between the resonances 1:1
and 1:3.
We apply Chirikov's criterion in order to have an estimate of
$\varepsilon_n$ [see Eq.(\ref{eq:chirikov})]~:
$$
\varepsilon_n^{(c)}=\frac{1}{4(n+1)^2}.
$$
For the largest one $n=1$, this value is equal to $0.0625$ which is
approximately twice the value obtained by renormalization. Escande's
approximate renormalization gives 0.0352 as the critical amplitude of
the field for the last invariant torus, which is close to the value
determined by the complete renormalization method. This feature
is expected to be true for quadratic Hamiltonians in the actions as it
was pointed out in Refs.~\cite{reic84,lin85}. In Ref.~\cite{chan99a}, 
it has been
noticed that the approximate renormalization
usually slightly overestimates the real critical value. \\
Lin and Reichl~\cite{lin86} have developed a method adapted to the
specific model~(\ref{eq:Ham1.5}) to compute the critical amplitudes
$\varepsilon_n$.  This method is based on the fact that as soon as the
last invariant torus is broken, some trajectories starting near one of
the main resonance (say 1:$n$) can approach the other main resonance
1:$n$+2. Since the diffusion of these trajectories can be very slow
(due in particular to the resonances of low order between two main
neighboring resonances, and in particular in the region of phase space
where the last KAM torus breaks up when the parameter $\varepsilon'$
is close to its critical value), the values they obtained overestimate the
ones obtained by renormalization. For instance, between 1:1 and 1:3
the critical value they obtained is approximately 0.037$\pm$0.001.\\
For $n=3$, the estimate obtained by Chirikov's criterion is
$\varepsilon_3^{(c)}\approx 0.156$, Escande-Doveil's renormalization
gives $\varepsilon_3\approx 0.0080$, whereas Lin and Reichl obtained
$\varepsilon_3\approx 0.0081\pm 0.0003$. The approximate value of
$\varepsilon_3$ obtained by the complete renormalization is
$\varepsilon_3\approx 0.0068$. Again, the same comments apply to this
case~: Lin and Reichl's method overestimates the value obtained by
renormalization because of the very slow diffusion of the trajectories
between the resonances 1:3 and 1:5 for critical or near-critical
values of the parameter $\varepsilon'$.

\subsection{Critical function $\varepsilon_c(\omega ; m,a,\Omega)$}
\label{sec:42}

We have studied the critical function $\varepsilon'_c(\omega)$ of
Hamiltonian~(\ref{eq:Ham-aa2}). This rescaled Hamiltonian is
equivalent to the initial Hamiltonian~(\ref{eq:Ham2}) with
$ma\Omega^2=1$ and in particular with $a=1$, $m=1$ and $\Omega=1$. For
other values of $a$, $m$, and $\Omega$, the critical function
$\varepsilon_c(\omega ; m,a,\Omega)$ is equal to
$$
\varepsilon_c(\omega ; m,a,\Omega)=ma\Omega^2
\varepsilon'_c\left(\frac{\omega}{\Omega}\right),
$$
according to Eq.~(\ref{eq:epsdl}). We notice that the argument of
$\varepsilon'_c$ is the {\it rescaled} frequency $\omega/\Omega$ since time
has been rescaled by a factor $\Omega$. 
Thus the critical function
varies like the square of the frequency of the field. In particular,
the largest value of the parameter for which an invariant torus
persists between the resonances 1:$n$ and 1:$n$+2 varies like~:
$$
\varepsilon_n(\Omega)=\Omega^2\varepsilon_n(\Omega=1),
$$
which is
consistent with the numerical results found in Ref.~\cite{lin86}. This
feature can be generalized to the following one-parameter family of
Hamiltonians~:
$$
H(p,x,t)=\frac{p^2}{2m}+V_{SQ}(x)+\varepsilon\sum_i
f_i(x)g_i(\Omega t),
$$
where $g_i$ are $2\pi$-periodic functions. This can be seen by
rescaling time by a factor $\Omega$
$$
H'(p,x,t')=\frac{1}{\Omega} H\left( p,x,\frac{t'}{\Omega}\right),
$$
and the actions by a factor $1/\Omega$
$$
H''(p',x,t')=\frac{1}{\Omega}H'(\Omega p',x,t').
$$
The fact that the critical coupling
$\varepsilon_c(\omega;m,a,\Omega)$ is proportional to $m$ is general
for a particle in an infinite square-well potential driven by a
perturbation depending on $x$ and periodically on $t$. This is
obtained by rescaling the momentum by a factor $1/m$ of the
Hamiltonian~:
$$
H(p,x,t)=\frac{p^2}{2m}+V_{SQ}(x)+\varepsilon V(x,t).
$$
From the fact that the interaction with the field is proportional
to $x$ (the interaction is of the form $\varepsilon x f(\Omega t)$
where $f$ is $2\pi$-periodic), the critical function
$\varepsilon_c(\omega;m,a,\Omega)$ is expected to be proportional to
$a$.

\section{Conclusion}

We have defined and studied numerically a renormalization
transformation for a system of a particle in an infinite square well
potential driven by an external monochromatic field in order to
determine the critical thresholds of the break-up of invariant tori.
Renormalization allows to obtain very precise information on the
stability of Hamiltonian systems with two degrees of freedom, and in
particular it allows to determine what kind of stable motions remains
as a function of the amplitude of the perturbation. We have checked
that for some specific frequencies, the thresholds obtained by
renormalization coincide with the ones obtained by Greene's residue
criterion.\\ 
We have chosen to
study a particle in an infinite square well potential driven by an external
field because of its simplicity, but the renormalization methods are
very general and can be applied directly to other types of Hamiltonian
systems with two effective degrees of freedom.

\section*{acknowledgments}

The author acknowledges useful discussions with G.\ Benfatto, A.\
Celletti, H.R.\ 
Jauslin, H.\ Koch, J.\ Laskar and R.S.\ MacKay.

\end{document}